\input harvmac
\def \la{\longrightarrow}
\def \up {\uparrow}
 \def \upa {\uparrow}
 \def \nea {\nearrow}

\def\ma{\mapsto}

\def \k {\kappa}

\def \del {\partial}

\def \ha{{\textstyle{1\over 2}}}

\def \a {\alpha}
\def \b {\beta}

\def \H {\td H} 

\def \s {\sigma}
\def \p {\phi}
\def \m {\mu}
\def \n {\nu}

\def \t {\theta}
\def \td {\tilde }

\def \K {{\tilde K}}

\def \inv {^{-1}}
\def \ov {\over }

\def \Q {{\cal Q}}

\def \lr { \lref}
\def\np {{  Nucl. Phys. }}
\def \pl {{  Phys. Lett. }}
\def \mpl {{ Mod. Phys. Lett. }}
\def \prl {{  Phys. Rev. Lett. }}
\def \pr  {{ Phys. Rev. }}

\def \cqg {{ Class. Quant. Grav. }}

\baselineskip8pt
\Title{
\vbox
{\baselineskip 6pt{\hbox{ CERN-TH/96-321}}{\hbox
{Imperial/TP/96-97/03 }}{\hbox{hep-th/9611047}} {\hbox{
  }}} }
{\vbox{\centerline { Waves, boosted branes  and  BPS 
states  in M-theory}
\vskip4pt
}}
\vskip -27 true pt
\centerline  {  J.G. Russo{\footnote {$^*$} {e-mail address:
jrusso@vxcern.cern.ch
 } }}
\smallskip \smallskip
\centerline{\it Theory Division, CERN, Geneva\   and  \   SISSA, Trieste  }
\medskip
\centerline {and}
\medskip
\centerline{   A.A. Tseytlin\footnote{$^{\star}$}{\baselineskip8pt
e-mail address: tseytlin@ic.ac.uk}\footnote{$^{\dagger}$}
{\baselineskip8pt On leave  from Lebedev  Physics
Institute, Moscow.} 
}

\smallskip
\centerline{{\it Theory Division, CERN, Geneva\   and  \  Blackett 
             Laboratory,  Imperial College,  London}}

\medskip
\centerline {\bf Abstract}
\medskip
\baselineskip10pt
\noindent
Certain type II string non-threshold 
BPS bound states  are shown to be related to 
non-static  backgrounds in 11-dimensional theory.
The 11-d counterpart of the bound state of NS-NS 
and R-R type IIB strings wound around a circle is a  pure gravitational 
wave propagating along a generic cycle of 2-torus. The extremal 
$(q_1,q_2)$ string with non-vanishing momentum along the circle (or 
infinitely boosted black string) corresponds in $D=11$ to a 2-brane 
wrapped around 2-torus with momentum flow along  the $(q_1,q_2)$ cycle. 
Applying duality transformations to the string-string solution we find 
the type IIA background representing the bound state of 2-brane and 
0-brane. Its lift to 11 dimensions is simply a 2-brane finitely boosted 
in  transverse direction. This 11-d solution interpolates between a 
static 2-brane (zero boost) and a gravitational wave in 11-th dimension 
(infinite boost). Similar interpretations are given for various bound 
states involving 5-branes. Relations between transversely boosted M-branes 
and 1/2 supersymmetric non-threshold bound states 2+0 and 5+0 complement 
those between M-branes with  momentum in longitudinal direction and 
1/4 supersymmetric threshold bound states 1+0 and 4+0. 
In the second part of the paper we  establish the correspondence 
between  BPS  states of  type IIB strings on a circle and oscillating 
states of a fundamental supermembrane wrapped around  2-torus. We show 
that the $(q_1,q_2)$ string spectrum is reproduced by the 
membrane BPS spectrum, determined using a certain limit. 
This supports the picture suggested  by Schwarz.

\medskip
\Date {November 1996}
\noblackbox
\baselineskip 14pt plus 2pt minus 2pt

\lr \dgh {A. Dabholkar, G.W. Gibbons, J. Harvey and F. Ruiz Ruiz,  \np
B340 (1990) 33;
A. Dabholkar and  J. Harvey,  \prl
63 (1989) 478.
}
\lr\mon{J.P. Gauntlett, J.A. Harvey and J.T. Liu, \np B409 (1993) 363.}
\lr\chs{C.G. Callan, J.A. Harvey and A. Strominger, 
\np { B359 } (1991)  611.}

\lr \CM{ C.G. Callan and  J.M.  Maldacena, \np B472 (1996) 571,  hep-th/9602043.} 
\lr\SV {A. Strominger and C. Vafa, HUTP-96-A002,  hep-th/9601029.}

\lr\MV {J.C. Breckenridge, R.C. Myers, A.W. Peet  and C. Vafa, HUTP-96-A005,  hep-th/9602065.}

\lr \CT{M. Cveti\v c and  A.A.  Tseytlin, 
\pl { B366} (1996) 95, hep-th/9510097. 
}
\lr \CTT{M. Cveti\v c and  A.A.  Tseytlin, 
IASSNS-HEP-95-102, hep-th/9512031. 
}
\lr\LW{ F. Larsen  and F. Wilczek, 
PUPT-1576,  hep-th/9511064.    }
\lr\TT{A.A. Tseytlin, \mpl A11 (1996) 689,   hep-th/9601177.}
\lr \HT{ G.T. Horowitz and A.A. Tseytlin,  \pr { D51} (1995) 
2896, hep-th/9409021.}
\lr\khu{R. Khuri, \np B387 (1992) 315; \pl B294 (1992) 325.}
\lr\CY{M. Cveti\v c and D. Youm,
 UPR-0672-T, hep-th/9507090; UPR-0675-T, hep-th/9508058; 
  \pl { B359} (1995) 87, 
hep-th/9507160.}

\lr\ght{G.W. Gibbons, G.T. Horowitz and P.K. Townsend, \cqg 12 (1995) 297,
hep-th/9410073.}
\lr\dul{M.J. Duff and J.X. Lu, \np B416 (1994) 301, hep-th/9306052. }
\lr\hst {G.T. Horowitz and A. Strominger, hep-th/9602051.}
\lr\dull{M.J. Duff and J.X. Lu, \pl B273 (1991) 409. }
\lr \guv{R. G\"uven, \pl B276 (1992) 49. }
\lr \gups {S.S. Gupser, I.R.   Klebanov  and A.W. Peet, 
hep-th/9602135.}
\lr \dus { M.J. Duff and  K.S. Stelle, \pl B253 (1991) 113.}

\lr\hos{G.T.~Horowitz and A.~Strominger, Nucl. Phys. { B360}
(1991) 197.}
\lr\teit{R. Nepomechi, \pr D31 (1985) 1921; C. Teitelboim, \pl B167 (1986) 69.}
\lr \duf { M.J. Duff, P.S. Howe, T. Inami and K.S. Stelle, 
\pl B191 (1987) 70. }
\lr\duh {A. Dabholkar and J.A. Harvey, \prl { 63} (1989) 478;
 A. Dabholkar, G.W.   Gibbons, J.A.   Harvey  and F. Ruiz-Ruiz,
\np { B340} (1990) 33. }
\lr\mina{M.J. Duff, J.T. Liu and R. Minasian, 
\np B452 (1995) 261, hep-th/9506126.}
\lr\dvv{R. Dijkgraaf, E. Verlinde and H. Verlinde, hep-th/9603126.}
\lr\gibb{G.W. Gibbons and P.K. Townsend, \prl  71
(1993) 3754, hep-th/9307049.}
\lr\town{P.K. Townsend, hep-th/9512062.}
\lr\kap{D. Kaplan and J. Michelson, hep-th/9510053.}
\lr\hult{
C.M. Hull and P.K. Townsend, Nucl. Phys. { B438} (1995) 109;
P.K. Townsend, Phys. Lett. {B350} (1995) 184;
E. Witten, \np B443 (1995) 85; 
J.H. Schwarz,  \pl B367 (1996) 97, hep-th/9510086, hep-th/9601077;
P.K. Townsend, hep-th/9507048;
M.J. Duff, J.T. Liu and R. Minasian, 
\np B452 (1995) 261, hep-th/9506126; 
K. Becker, M. Becker and A. Strominger, Nucl. Phys. { B456} (1995) 130;
I. Bars and S. Yankielowicz, hep-th/9511098;
P. Ho{\v r}ava and E. Witten, Nucl. Phys. { B460} (1996) 506;
E. Witten, hep-th/9512219.}
\lr\beck{
K. Becker and  M. Becker, hep-th/9602071.}
\lr\aar{
O. Aharony, J. Sonnenschein and S. Yankielowicz, hep-th/9603009.}
\lr\ald{F. Aldabe, hep-th/9603183.}
\lr\ast{A. Strominger, hep-th/9512059.}
\lr \ttt{P.K. Townsend, hep-th/9512062.}
\lr \papd{G. Papadopoulos and P.K. Townsend, \pl B380 (1996) 273, hep-th/9603087.}
\lr\jch {J. Polchinski, S. Chaudhuri and C.V. Johnson, 
hep-th/9602052.}
\lr \ddd{E. Witten, hep-th/9510135;
M. Bershadsky, C. Vafa and V. Sadov, hep-th/9510225;
A. Sen, hep-th/9510229, hep-th/9511026;
C. Vafa, hep-th/9511088;
M. Douglas, hep-th/9512077. }

\lr \duflu { M.J. Duff and J.X. Lu, \np B354 (1991) 141. } 
\lr \pol { J. Polchinski, \prl 75 (1995) 4724,  hep-th/9510017.} 
\lr \iz { J.M. Izquierdo, N.D. Lambert, G. Papadopoulos and 
P.K. Townsend,  \np B460 (1996) 560, hep-th/9508177. }

\lr \US{M. Cveti\v c and  A.A.  Tseytlin, 
\pl {B366} (1996) 95, hep-th/9510097;   hep-th/9512031.  
}

\lr \green{M.B. Green and M. Gutperle, hep-th/9604091.}

\lr \berg{E. Bergshoeff, C. Hull and T. Ort\' \i n, \np B451 (1995) 547, hep-th/9504081.}

\lr \gibbon{G.W. Gibbons, \np B207 (1982) 337. }
\lr \hullo{C.M. Hull, \pl B139 (1984) 39. }
\lr \horts{G.T. Horowitz and A.A.  Tseytlin, \pr D51 (1994) 3351,
hep-th/9408040.}
\lr \horow{  J.H. Horne, G.T. Horowitz and 
 A.R. Steif, \prl 68 (1992) 568. }
\lr \dabwal {A. Dabholkar, J.P. Gauntlett, J.A. Harvey  and
 D. Waldram, \np B474 (1996) 85,  
hep-th/9511053. 
 }
\lr \tset  {A.A.  Tseytlin,  \np B475 (1996) 179, hep-th/9604035.}
\lr \john {J.H.  Schwarz, \pl B360 (1995) 13 (E: B364 (1995) 252),
hep-th/9508143, hep-th/9509148.}
\lr \johnt {J.H.  Schwarz, \pl B367 (1996) 97, 
 hep-th/9510086. }

\lr \townelev{ P.K. Townsend, \pl B350 (1995) 184, hep-th/9501068.  }
\lr \calmalpeet {C.G. Callan, J.M.  Maldacena and A.W. Peet, 
\np B475 (1996) 645,  hep-th/9510134. }
\lr \klts{I.R. Klebanov and A.A. Tseytlin,
 \np B475 (1996) 179,
hep-th/9604166. }
\lr \cvets{ M. Cveti\v c  and A.A. Tseytlin, \np B478 (1996) 181, 
hep-th/9606033. }
\lr \pope {N. Khvengia, Z. Khvengia, H. L\"u and C.N. Pope,
hep-th/9605077. }
 
\lr\paptkk{
G. Papadopoulos  and P.K. Townsend, hep-th/9609095. }

\lr\papadop{
G. Papadopoulos, hep-th/9604068. }

\lr \witten {E. Witten, \np B460 (1995) 335, hep-th/9510135.}

\lr \tow {P.K. Townsend,  hep-th/9609217. } 

\lr \schwa  {J.H. Schwarz,
hep-th/9607201.  }
\lr \duff { M.J. Duff, hep-th/9608117.  }
\lr \bergsh{ E. Bergshoeff, E.  Sezgin and P.K. Townsend, \pl B189 (1987)
75.}
\lr \doug {M.R. Douglas,  hep-th/9512077.}
\lr \gaunt {J. Gauntlett, D. Kastor and J. Traschen, hep-th/9604189.}
\lr \aspin {P. Aspinwall,  hep-th/9508154.   }
\lr \lifsh {G. Lifschytz, hep-th/9610125. }

\lr\grepap{
M.B. Green, N.D. Lambert, G. Papadopoulos  and P.K. Townsend, 
hep-th/9605146. }

\lr\banks{
T. Banks, W. Fischler, S.H. Shenker and L. Susskind, 
   hep-th/9610043.}

\lr \tser{A.A. Tseytlin, hep-th/9609212. }

\lr\costa { M.S. Costa, hep-th/9609181.}

\lr \schmid {C.  Schmidhuber, \np B467 (1996) 146,   hep-th/9601003.  }
\lr \tsetli{ A.A. Tseytlin, \np B469 (1996) 51, hep-th/9602064.  }

\lr\russo {J.G. Russo, 
hep-th/9610018 .}
\lr\dewitt { B. de Wit, J. Hoppe and H. Nicolai, 
\np  {B305} [FS 23] (1988) 545.}
\lr\dufi{ M. Duff, T. Inami, C. Pope, E. Sezgin and K. Stelle,
\np {B297} (1988) 515.}
\lr\bst{E. Bergshoeff,  E. Sezgin and Y. Tanii, \np {B298}
(1988) 187.}
\lr \bergsh { E. Bergshoeff,  E. Sezgin and P.K. Townsend,
\pl { B189} (1987) 75;
 Ann. Phys. {185} (1988) 330.} 
\lr\dewit { B. de Wit, M. L\" uscher and H. Nicolai, 
\np  {B320} (1989) 135.}

 \lr\dddo{M.R. Douglas, D. Kabat, P. Pouliot and  S.H. Shenker,
  hep-th/9608024. }

\lr\beh {K. Behrndt, E. Bergshoeff and B. Janssen, hep-th/9604168
(revised).}
\lr\verl{E. Verlinde, \np B445 (1995) 211.}
\lr\ber{M. Bershadsky, C. Vafa and V. Sadov, \np B463 (1996) 398, 
hep-th/9510225.}
\lr\sen{A. Sen, \pr D53 (1996) 2964, 
hep-th/9510229;   \pr D53 (1996)  2874,  hep-th/9511026.}

\lr \give{A. Giveon and M. Porrati, hep-th/9605118.}

\lr\kaa{ E.A. Bergshoeff, R. Kallosh and T. Ort\'\i n, 
    \pr D47 (1993) 5444.}
\newsec{Introduction}

In view of  recent suggestions that  $D=11$ supergravity 
may be a  low-energy effective field theory of  
a more fundamental  `M-theory', 
it is  important to clarify further how 
different p-branes and their  BPS bound state  
configurations in $D=10$ string theories 
can be understood from  eleven-dimensional perspective
 (for  recent reviews and refs. see 
\refs{\schwa,\duff,\tow}). 
At the level of   classical configurations of the effective field theory, 
the  threshold  (i.e. zero binding energy)
BPS  bound states of p-branes 
(corresponding, in particular, to superpositions  
 of D-branes \refs{\witten,\ber,\sen}  with $p-p'=4,8$
\refs{\jch\doug\green -\dddo}) 
originate from combinations of 
intersecting 2-branes and 5-branes in 11
dimensions \refs{\papd\tset\klts\gaunt
\pope -\paptkk}. 

As for   non-threshold  BPS bound states 
with non-zero binding energy (typically, with  $M_{1+2}
 =\sqrt {M_1^2 + M_2^2}$)
which  cannot be viewed as  $n$ weakly coupled p-branes 
at equilibrium (and  
are not described  by $n$ independent 
 harmonic functions),  
 their 11-dimensional origin  was not much discussed in the past.
 The basic example  of such $D=10$ configuration is 
 the $(q_1,q_2)$  bound state of the NS-NS and R-R strings 
in type IIB theory 
\refs{\john,\johnt,\schwa,\witten}, from which various  
other  similar  bound states 
(corresponding, in particular, to 
superpositions of D-branes with $p-p' = 2,6$ \refs{\jch,\green,
\lifsh})  may be constructed 
by applying  $T$ and $SL(2,Z)$ dualities.

It was 
suggested  in \refs{\john,\schwa}  from the  consideration 
of the  nine-dimensional 0-brane spectra that 
the  $(q_1,q_2)$  string-string  type IIB bound states 
  should be related to  the  states of 
 2-brane wrapped around  a 2-torus in $D=11$.
 One of the aims of the present paper 
 is to extend and complete  what was done in  \refs{\john,\schwa}
 at two different levels.  
 At the `macroscopic'  classical-solution level, 
 we shall explicitly  identify the $D=11$ background which has the
 $(q_1,q_2)$  string solution of \john\ 
  as  $T$-dual of its  dimensional reduction.
 We shall also determine the $D=11$ counterparts of  several
 type IIA  non-threshold bound state backgrounds 
  related to the string  one by  $T$ and $SL(2,Z)$ dualities. 
 At the  microscopic   quantum-state level, 
 we shall  demonstrate that 
 the mass spectrum  of   BPS 
states of $(q_1,q_2)$  string   indeed  matches  the 
 spectrum  of the corresponding 
 oscillating  states of  wrapped membrane  
 not only for the  zero-mode parts of the masses,   
 as was already  shown  in \refs{\john,\schwa}, 
  but also including the  oscillator parts.

 Section 2 will be devoted to a  discussion
 of   classical type II $D=10$
 solutions corresponding to various 
 non-threshold BPS bound states of branes and their $D=11$ 
 counterparts.
 We shall first determine (in Section 2.1) which  
   $D=11$  solution  is related, by dimensional reduction and
$T$-duality,    to the 
$(q_1,q_2)$ string  type IIB  background
constructed  in \john\
by applying $SL(2,R)$ duality transformation to the fundamental 
$(1,0)$ string solution of \dgh.
In the general case of  the  $(q_1,q_2)$ string with momentum 
(which is a straightforward generalization of the solution 
of \john\ and preserves  1/4 of maximal supersymmetry) 
the corresponding  $D=11$ solution is the extremal  limit of a 
2-brane \refs{\dus}  wrapped around a 2-torus 
and  infinitely `boosted' \refs{\tset,\cvets} 
along  {\it generic}  $(q_1,q_2)$ cycle of the torus.
In agreement with the 
 interpretation  given  in \refs{\john},
  the   momentum (or the  charge 
of the `boost' harmonic function) 
of the $(q_1,q_2)$ string
is  the  winding number (charge) of the 2-brane, 
 while the winding number of the $(q_1,q_2)$ string
 is  the  momentum of  the 2-brane boosted along the 
 $(q_1,q_2)$ cycle. The counterpart of the zero-momentum
 string is thus a zero-charge limit of 2-brane boosted to the
 speed of light, or simply a 
 gravitational wave along the cycle of 2-torus.
 
 For arbitrary  values of the charges, this 
  1/4 supersymmetric  $D=11$ solution
   can be thought of as a bound state of a 2-brane and a
 gravitational wave (or $2+ \nearrow$, where the direction of
 the arrow indicates  a `diagonal'  direction of the momentum flow
 on 2-torus).  In analogy with the fundamental string case 
 \refs{\dgh,\calmalpeet,\dabwal}, 
 it should  represent, at macroscopic level, 
  an   excited  BPS state  of  the  membrane 
   with the momentum being carried by 
    oscillations propagating along 
  the $(q_1,q_2)$ cycle of the torus. 
Its dimensional reduction
 is  type IIA solution  
  representing  a bound state of a fundamental string and a 
  0-brane with a boost along the string, $ 1 + 0\ma $.
 The $T$-duality  in  the string  direction 
 relates this to  our  starting point,  type IIB bound state 
 of the wave, R-R string and 
 fundamental string, $\upa+ 1_{R} + 1_{NS}$.

In general,   bound states of type IIB branes, 
constructed  by applying $SL(2,R)$ duality transformations,
can be also obtained  from  the 11-dimensional  theory by  starting
 with  an   appropriate 
 M-brane configurations   and boosting them  along a 
non-trivial $(q_1,q_2)$ cycle of the 2-torus. This is 
in agreement with 
the  observation
\refs{\john,\aspin} that $SL(2,Z)$ duality of type IIB theory has its
origin in the  toroidal $SL(2,Z)$ in $D=11$
(in the case of $D=11$ backgrounds with at least  two isometries which we
shall be considering  here this  was originally  understood   in \berg).
We shall  illustrate this  further in Section 2.2   on several 
  examples,  starting with  
the $(q_1,q_2)$ bound state of NS-NS and R-R 5-branes
\refs{\john,\witten}.  The   1/4 supersymmetric 
$1_{NS}  + 1_R + \up $ and $5_{NS}  + 5_R + \up$
configurations  are the special cases 
  of the  non-threshold 1/8 supersymmetric 
configuration 
$ (1 + 5)_{NS}  + (1+ 5)_{R} +\up$ which is an   
$SL(2,Z)$ rotation of the threshold bound state of the NS-NS string and 
solitonic 5-brane \TT\  or of its $SL(2,Z)$ dual R-R version \CM.
Its $D=11$ counterpart is shown to be  the intersection 
of the 2-brane and 5-brane over a string \tset\  with  a momentum flow 
along  generic `diagonal' direction of the 2-brane  torus, i.e.
$5\bot2 + \nea$.

In Section 2.3 we shall address the question of  
$D=11$ interpretation of other non-threshold  type II BPS bound states
related to the  $(q_1,q_2)$ string by the  duality transformations. 
We shall demonstrate that
 the  type IIA 
bound state of a  2-brane and a 0-brane  is  just a dimensional reduction 
of the standard 
 extremal $D=11$ 2-brane solution {\it finitely} ($v <1$) 
 boosted  along the  isometric 
11-th dimension  which is {\it transverse} to the brane, or
 $2 \ma$. 
Analogous surprisingly simple  interpretation
  (which  is  not  obvious 
  at the level of the type IIA backgrounds) 
 applies also
to the $5+0$  bound state.
The   relations 
 between  boosted 
 branes in $D=11$ and   `p-brane + 0-brane'  bound states in $D=10$
 can be summarized as follows:
 (a) 1/4 supersymmetric threshold  bound states:
 $ 2 + \up\ \la  1+0 , \ \ 5 + \up\ \la  4+0 $; \  
 (b) 1/2 supersymmetric non-threshold  bound states:
 $ 2 \ma\ \la  2+0 , \ \ 5 \ma\ \la  5+0 $. 
We shall   discuss  two  natural generalizations, $2\bot 2\ma$ and 
$5\bot 2 \ma$.
In particular, $2\bot1 +0$   solution  ($T$-dual 
to $1_{NS} + 1_R + \up$)  is the reduction of the  $D=11$ 
 configuration  
$2 \bot 2 \ma $  in which one of the two 2-branes which  intersect
over a point \papd\ is finitely boosted in a 
direction of  the other 2-brane  orthogonal to it.  
 We shall also comment on 
 the existence of several different 11-dimensional solutions
 which reduce to $D=10$ solutions related by $T$ and $SL(2,Z)$ dualities.

 In  Section 3  we shall address the  relation  between 
 the    type IIB strings and 
  $D=11$   membrane at the `microscopic'
 level,  by
 explicitly identifying the  quantum  supermembrane  states 
  which correspond to the BPS spectrum of the bound states 
  of type IIB  strings.  These  excited membrane states  
  have oscillations only along one  (momentum)  direction, 
  in agreement with  what is also implied by 
  the macroscopic  effective field theory picture. 
 As was shown in 
   \refs{\john,\schwa},  the zero-mode part of  the mass 
of   BPS states of type IIB  $(q_1,q_2)$ string on a circle 
is in perfect agreement 
 with the zero-mode part of the mass of 
  the fundamental supermembrane  states wrapped around 
a  2-torus with a momentum along the $(q_1,q_2)$ cycle of the 
torus. Our aim will be to  check  that the oscillator parts in 
the masses  also agree. 
 We shall argue  that in order to determine the
 relevant  BPS spectrum of the supermembrane  is sufficient to  
 calculate it in a certain region of the coupling parameters.
Solving the resulting gaussian theory,    we shall  find
 that  its  oscillating membrane BPS  states 
 do, indeed, have the same  masses as the BPS states
 of type IIB strings.

\newsec{Boosted branes  and  non-threshold BPS bound
states: classical solutions}
Gravitational waves  carrying linear  momentum 
 seem to play as important  role
 in  11-dimensional  theory as  2-brane and 5-brane solutions.  
This may not be unexpected  since  plane fronted 
  waves are known to preserve supersymmetry
  \refs{\gibbon,\hullo,\kaa};    they     
are related, in $D=10$ theory,  to the
  fundamental string backgrounds by
$T$-duality \refs{\horow,\horts};   and   
a single plane fronted wave (Schwarzschild background
 boosted to the speed of
light)
is the  $D=11$ image of  the  0-brane of $D=10$ 
type IIA theory
\refs{\gibbon,\townelev}. 
Nevertheless, it is a bit surprising  (though, in fact,
 implicit in \john) 
that the $D=11$ solution 
that corresponds to  a rather complicated-looking 
$(q_1,q_2)$ string background of type IIB theory 
turns out to be  related to  a dimensional reduction of the 
 pure gravitational wave  propagating along 
the $(q_1,q_2)$ cycle of 2-torus in $M^{11}= T^2 \times M^9$
 which  describes 
 a zero-charge limit of 2-brane boosted to the speed of light.
Wrapping  a  (non-zero charge)  2-brane     
around $T^2$ corresponds to adding 
a  momentum (wave)  to the type IIB $(q_1,q_2)$ string \john.
As in the case of the $D=10$ fundamental string 
 \refs{\dabwal,\calmalpeet},  
here the wave may be represented  by 
transverse oscillations
of the 2-brane propagating  along the cycle (i.e. 
in {\it one} circular 
 direction)  and carrying 
the corresponding  momentum. 
This suggests that, as in the  string-theory
case,  there should exist  the 
corresponding BPS states of the quantum
supermembrane. This issue will be addressed in Section 3, 
while below  we shall  discuss only the classical 
effective field theory solutions.

\subsec{\bf Wave along generic cycle of  2-torus  in  $D=11$
 and  $SL(2,Z)$   family 
of type IIB strings}
In what follows we shall  consider the 11-dimensional space 
$M^{11}= T^2 \times M^9$  with the isometric
rectangular 2-torus   coordinates  $(y_1,y_2)$ 
having periods $(2\pi R_1, 2\pi R_2)$ (the case of more general 
$T^2$ is treated similarly). 
The gravitational wave propagating in $y_1$ direction ($i=1,...,8$) 
\eqn\wav{
ds^2_{11} = dudv + W(u,x) du^2 + dy_2^2  + dx_i dx_i \ ,  \ 
\ \   \del^2_x W=0 \ ,  \ \ \  u,v= y_1 \pm t \ ,   }
solves the  $D=11$ Einstein equations for any harmonic  function $W$. 
The case of $W= { \Q\ov |x|^6} $ is special as this pp-wave 
can be represented  \gibbon\ as the  extremal
(infinite boost, zero mass) limit of the 
boosted  Schwarzschild solution, or, more precisely, since $y_1, y_2$ are
isometries, of  a black 2-brane of  zero charge, 
\eqn\wave{
ds^2_{11} =  -f(r) dt'^2  + dy'^2_1  + dy_2^2  + 
f\inv (r) dr^2 +  r^2 d\Omega_7^2  }
$$ 
=  - dt^2 + dy^2_1  + dy^2_2  + 
 {\mu\over r^6} (\cosh \beta\ dt - \sinh \beta\ dy_1)^2
 + f\inv (r) dr^2 +  r^2 d\Omega_7^2  $$
 $$ =  - \hat K^{-1} (r) f(r) dt^2 +  \hat K(r) [dy_1+ A (r) dt]^2 \  + dy_2^2   
 + f\inv (r) dr^2 +  r^2 d\Omega_7^2  \ , $$
 where 
 $$ 
 \ \  t'= \cosh \beta \ t - \sinh \beta\ y_1\ ,  \ \ \ \ \ 
\ y'_1= - \sinh \b\ t + \cosh \beta\ y _1 \ , $$
\eqn\www{ f= 1 - {\mu\over r^{6}} \ , \ \ \  \hat K = 1 + { \hat \Q \over r^{6}} \ , 
\ \ \ \   A = - { \Q \ov r^6}  \hat K \inv \ , }
$$
   \hat \Q = \mu \sinh^2 \beta \ , \ \ \ \  \ \
\Q = \mu  \sinh \beta \cosh \beta\ . $$
The 
 boost parameter  $\b$  is related to the  Kaluza-Klein electric 
charge   $\Q$, i.e. the momentum along the boost 
direction $y_1$.
In the extremal  limit 
$\mu\to 0, \ \beta\to \infty,  \ \Q=$fixed,    
  the metric \wave\ takes the form \wav\ ($\hat \Q=\Q,\ 
 K={\hat K}$)
with $W= K-1 = {\Q\ov r^6}$.

The 1/2 supersymmetric  wave  can be superposed 
with 1/2 supersymmetric extremal 2-brane 
(this can be shown, e.g.,  by boosting the black 2-brane \guv\ and taking the extremal
limit). The  resulting  1/4 
supersymmetric  background is 
\refs{\tset,\cvets}
 $$
d s^2_{11} = H^{1/3}_2  (r) 
 \big( H\inv _2 (r)  \left[-    dt^2  + dy^2_1 +  dy^2_2 + 
   W(r) (dt - dy_1)^2 \right] +   dx_i dx_i   \big) \ , 
      $$ \eqn\mem{
 C_3  =  H_2\inv dt\wedge   dy_1\wedge d y_2 \  , \ \ 
\ \ \   H_2= 1 + {Q \over r^6}\ , \ \ \   W= {\Q\ov r^6 } \ . }
More generally, $W(r)$  in \mem\ 
can be replaced by any solution  $W(u,x)$
of the Laplace equation $\del^2_x W=0$. 
In particular, as in the string case 
\refs{\dabwal, \calmalpeet}, choosing 
 $W= f_i (u) x^i$  corresponds to adding a 
wave  of  membrane oscillations propagating along $y_1$ ($f_i$  is 
related to the profile of oscillations). The macroscopic properties of such
solution are the same as of the solution with $ W= {\Q\ov r^6 }$
with $ \Q$ being proportional to the asymptotic value of 
momentum carried by the wave
($\sim \langle [ \int du f_i(u)]^2 \rangle $).

It is important to
stress  that one is able to construct  a supersymmetric  BPS 
background  by making an infinite  boost  in {\it one} isometric 
direction only
(boosting in $y_1$ and $y_2$ directions with 
two independent parameters $\b_a$ and 
taking the limit $\m\to 0, \b_a \to \infty$ 
does not  give  a new extremal solution). 
 Similar observation applies 
  in  the general p-brane case and suggests that the BPS states
of p-branes are always  `string-like', i.e.
have oscillations
in {\it one} 
 direction only.
  This should also follow
directly from the  BPS requirement  
and  supersymmetry algebra (cf. \dvv).

Dimensionally reducing the  `2-brane  plus wave' (or  $2 + \upa$) 
solution \mem\  to $D=10$  along the  `spectator' direction $y_2$ 
 we get  the `fundamental string  plus wave' (or $1 + \upa$)
  background.  Under the $T$-duality in $y_1$, it
 transforms into
the same  type  of solution of type IIB theory 
$1_{NS} + \upa\ $ 
(with $ H_2 \leftrightarrow K=1+W$, i.e. $Q \leftrightarrow  \Q$). 
If instead we reduce  along the boost direction $y_1$, 
  we get the type IIA solution, which can be interpreted 
  as a bound state of the fundamental string and the 0-brane, $1 + 0$.
  The $T$-dual solution of type IIB theory is 
  the boosted R-R string, 
  or $1_{R} + \upa.$ 
  This  suggests  that reducing to $D=10$ along  some `mixed' 
  direction  we should   get a more general class of solutions 
  which will interpolate
   between the two  special cases $ 1 + \upa$ and 
  $ 1 + 0$, or, when  $T$-duality transformed into
   type IIB theory,  
  between $1_{NS} + \upa$ and   $1_{R} +
  \upa$. 
   This is equivalent to putting  the  wave (i.e.  momentum)  
  along  some   generic  cycle   of
   the 2-torus. 
  
   For the simple rectangular torus $ y_a = R \s_a, \ \tau=i$, \ 
$ds^2 = R^2  |d\s_1 + \tau d\s_2|^2$, \ $\s_a \in (0,2\pi),$
 the direction of the $(q_1,q_2)$
 cycle\foot{The integers
$q_1,q_2$  can  be taken to be  relatively  prime since 
rescaling them by  common factor does not 
change the direction of the
cycle.}
is specified by 
 \eqn\sini{  \cos \t = { q_1 \ov \sqrt { q_1^2 + q_2^2}} \ , 
\ \ \ \ \sin  \t = { q_2 \ov \sqrt { q_1^2 + q_2^2} }\ .  } 
  The background corresponding to the 2-brane 
  boosted along the $(q_1,q_2)$ cycle  is then 
  found from \mem\ by a   rotation
 \eqn\rott{ y'_1  =  \cos \t \ y_1 + \sin \t\ y_2 \  ,  \ \ \ \ 
 \  y'_2 =  -\sin\t \ y_1 + \cos \t\ y_2 \ ,  }
 i.e.\foot{This 
background has, of course, 
 a straightforward non-extremal generalization, cf. \cvets.} 
 $$ 
d s^2_{11} = H^{1/3}_2  
 \left( H_2\inv [-    dt^2  + dy^2_1 +  dy^2_2 + 
   W (dt - \cos \t\ dy_1 - \sin \t\ dy_2 )^2 ] +  
    dx_i dx_i   \right), 
    $$ 
    \eqn\memo{  C_3  =  H_2\inv dt\wedge 
      dy_1\wedge d y_2 \ 
     .   } 
The quantized momentum has components 
$(nq_1,nq_2)$ and the modulus $n \sqrt {q_1^2 + q_2^2}$, 
implying the change 
 in  quantization condition for the
  Kaluza-Klein  charge  $\Q$
 \eqn\neww{
 \Q= c_0 {n \ov R}  \ \ \la \ \  \Q_q = c_0 {n\ov  R_q
 }   = \Q  \sqrt { q_1^2 + q_2^2}  \ ,   }
 where  $c_0 ={1 \ov 3} \k^2_9/\omega_7$, \ $\omega_7 = {1 \ov 3}\pi^4$, \    
 and $n$ is an integer.
 
Directing the boost along  the $(q_1,q_2)$  cycle 
of the torus restores the symmetry between $y_1$ and $y_2$ directions:
the background \memo\  is invariant under $y_1 \leftrightarrow y_2, 
\ q_1 \leftrightarrow q_2$, 
so that reductions along $y_1$ or $y_2$ give, for generic $(q_1,q_2)$, 
similar 
$D=10$ type IIA backgrounds
 related simply
by $q_1 \leftrightarrow q_2$.

Let us now show that the $D=11$ counterpart of 
 the  original  zero-momentum 
 $(q_1,q_2)$  string solution
 of  type IIB theory \john\
 is the limit of \memo\ in which 
  the 2-brane charge is set equal to zero,\foot{The non-extremal limit of this
  solution is a finitely  boosted black  (Schwarzschild) 
  2-brane  with  zero charge.}  
 $Q=0, \ H_2=1$, i.e.  
  a    zero-charge membrane boosted to the speed of light,
   or simply a 
  gravitational wave travelling along a cycle of $T^2$
  (in complete analogy with boosted string case \refs{\gibbon,
  \horow}.

  Rewriting the resulting metric as
$$ d s^2_{11} = -    dt^2  + dy^2_1 +  dy^2_2 + 
   W (dt - \cos \t\ dy_1 - \sin \t\ dy_2 )^2  +  
    dx_i dx_i 
$$
   \eqn\memoo{= \ - K^{-1} dt^2  + 
     K \K^{-1} [ dy_1 -   \cos \t \  W  K^{-1}  dt]^2 }
     $$
             +\  \K [ dy_2  -  \sin \t\ W   \K^{-1} dt
                           + \sin \t  \cos \t\  W \K^{-1}
                             dy_1 ]^2  + dx_i dx_i \ , $$
where
\eqn\defo{
K\equiv  1 + W \ , \ \ \ \   \K \equiv  1 +  {\rm sin}^2 \t\ W  \ ,  \ \ \ \ W= 
{ \Q_{q}\ov r^6 } \ ,  } 
one can easily read off the  type IIA solution 
which follows upon  the dimensional reduction along $y_2\equiv y_{11}$ 
direction.

The resulting  1/2 supersymmetric  type IIA background 
describes 
just a 0-brane {\it finitely}   boosted to a
 velocity  $v= \cos \t \leq  1 $ 
 in the isometric    direction $y_1$ (we shall denote this 
 configuration as $0 \ma$)
 \eqn\heq{
ds^2_{10A} =  \td K^{1/2} \big(  -  \td K\inv  d\td t^2 + 
d\td y_1^2  +  dx_i dx_i \big)  \ , }
$$ dA= d\td K\inv \wedge d\td t \ , \ \ \ \ 
e^{2\p} = \td K^{3/2}  \ , $$
where 
\eqn\hew{ \td t\equiv {1\ov \sin \t} ( t - \cos \t\ y_{1}) \ , \ \ \ \ 
\td y_{1} \equiv  {1\ov \sin \t} ( y_{1} - \cos \t\  t ) 
 \ . 
}
Note that $-\td t^2 + \td y^2_{1}  = - t^2 + y^2_{1}$
and also that the coefficient of the boosted 
0-brane  harmonic function $\tilde K$ 
depends on the boost parameter. 
 When  $q_2=0$  (infinite boost limit, 
 $\sin \theta \to 0$) 
  we get just a plane gravitational wave 
 in $y_1$ direction;  
 when  $q_1=0$  we  get 
a static 
 0-brane in the space with one compact isometric 
 direction,  or $0_1$. 
 
 $T$-duality along $y_1$ converts the boosted 0-brane into
 the string-string bound state of \john. 
Indeed, the R-R vector field originating from off-diagonal $(11,\m)$ components
of the metric  has both time-like 
and internal spatial parts,   which  transform    under  
 $T$-duality   into  the ${t y_1}$ component
 of the R-R  2-tensor $B^{(2)}_{\m\n}$  and the R-R scalar $\chi$ of type
 IIB theory.
Using the $T$-duality transformation rules,  
or directly,  the relation between a $D=11$ background with two isometries and 
the corresponding type IIB background 
 given in  \berg\
we find exactly  the  type IIB  $(q_1,q_2)$ string 
solution  obtained
 in \john\
by applying the $SL(2,R)$ transformation to the fundamental 
string 
 of \dgh\foot{We shall always use  string-frame  metric in $D=10$. Here $y_1$  is the 
  coordinate dual to $y_1$  in \memoo.}
\eqn\johhn{
ds^2_{10 B  } =   \K^{1/2} [ K\inv ( -dt^2 + dy^2_1) + dx_i dx_i ]\ , }
$$ e^{2\p}=   K\inv  \K^2  \ , \ \ \ \  \ \   \chi =  \sin \t  
 \cos \t\  W \K^{-1}  \ ,   $$ $$
 B^{(1)}_{ty_1}  =  - \cos \t\ W  K^{-1}  \ , \ \  \ \ \ \ 
  B^{(2)}_{t y_1}  =  - \sin \t \ W  K^{-1} \ .  $$
  Note that 
  $dB^{(1)} + i dB^{(2)} = e^{i\t} d K\inv \wedge dt\wedge dy_1$. 
  The special cases are the NS-NS string ($q_1=1, q_2=0$, i.e.
  $\t=0, \  \K=1$), and the R-R string 
  ($q_1=0, q_2=1$, i.e.
  $\t={\pi \ov 2} ,   \  \K=K$). 
  This is the background in the case of the   simplest 
  2-torus $\tau=i$ or to the trivial vacuum
  $ \rho_0 = \chi_0 + i e^{-\p_0} =i$  \john.
  The total momentum of the wave is just 
  the tension of the $(q_1,q_2)$ string.
  Generalization of the correspondence between
   \memoo\ and \johhn\  to the case of an arbitrary 
   2-torus is straightforward.
   In particular,  in 
  the case of the rectangular torus with
  different radii of $y_1$ and $y_2$, i.e. 
  $\tau_2 = R_1/R_2 $,  one  should   change 
  $q_2 \to {R_1\ov R_2} q_2$  in the above expressions 
  and thus replace  
  the $(q_1,q_2)$
  string tension factor (in the string frame) 
  $ \sqrt { q_1^2 + q_2^2}$ by 
  $ \sqrt { q_1^2 + e^{-2\p_0} q_2^2} =  
  \sqrt { q_1^2 +  ({R_1 \ov R_2} )^2  q_2^2} $  \john.
  

 All works in a similar  way if we add back the 2-brane, 
  i.e. switch on its charge,  
  starting  with \memo\ with  $H_2\not=1$,  and 
  reduce down to $D=10$ along
  $y_2$. The $D=11$ solution \memo\ then 
  leads to a 
  $D=10$ type IIA background which can be 
  interpreted as an effective  
  field theory representation of a 
  1/4 supersymmetric non-threshold
  bound state 
  of a fundamental string and a 0-brane boosted along the string
  direction, i.e. $ 1 + 0\ma$.
  $T$-duality   along $y_1$ transforms  it into the boosted
   version  of the  type IIB 
  $(q_1,q_2)$  string  which differs from \johhn\ only 
  by an extra  
 wave term  in the metric 
  \eqn\jhn{
ds^2_{10 B} =   \K^{1/2} \left( K\inv \left[ -dt^2 + dy^2_1 +  
 (H_2-1) (d t- d y_1)^2 \right] + dx_i dx_i \right)\ . }
  The 2-brane charge $Q$ (equal to $c_0 {w_0 R\ov \a' }$, where $w_0$
  is an  integer winding number)   thus becomes the momentum of the 
  $(q_1,q_2)$  string. 
   As in the case of  the fundamental NS-NS string
  \refs{\dabwal,\calmalpeet},
   the momentum flow  along $(q_1,q_2)$ string
   has also  a 
  microscopic representation  in terms of right-moving
  string oscillations, i.e. there exists a similar 
  solution 
   which is supported by an oscillating  string  source
   and has the same asymptotic value of momentum.
   
  This provides  an  explicit 
  classical-solution  realization of the observations
  in \john\ which were based on 
    the  expected form of the $D=9$  0-brane spectrum.  
  
\subsec{\bf Type IIB bound states of  strings  and 5-branes
  from $2\bot 5$ in $D=11$}

 Analogous  considerations 
apply in  general  to  type IIB 
solutions   obtained  by  making an  $SL(2,Z)$ transformation
  on    a `generating' background
  with  one or more compact isometries.
Since according to \refs{\berg,\john}
the action of $SL(2,Z)$ should correspond to a 
modular  transformation of the compactification  2-torus in $D=11$, 
such $SL(2,Z)$ families of solutions 
should  originate 
from  the $D=11$ counterpart of the generating solution
that is boosted  
 along  generic cycle of the  2-torus.\foot{This is somewhat similar to the
 description  of dyonic configurations
 in \refs{\verl,\give} and,   especially, in  \grepap,  where, in
 particular,  dyonic 2-branes in $D=8$ were 
 related  to  wrappings of self-dual  3-brane around generic cycle of
 2-torus.}

Let us  consider the  $SL(2,Z)$ 
bound states of NS-NS and R-R 5-branes in type IIB theory
\refs{\john,\witten} and explain their 11-dimensional origin.
 To  get  an idea of what kind of $D=11$ backgrounds we should
 expect to find, it is useful  to consider the  
 special cases, describing 
a type IIB solution, its possible  $T$-dual images in type IIA 
theory,  and $D=11$ solutions from which the latter 
 follow upon dimensional reduction:

(1)  solitonic NS-NS 5-brane of IIB theory 
with  a momentum flow along one of its directions 
$5_{NS} + \upa$  is $T$-dual  to: (a) bound state of 5-brane
and a wave   $5 + \upa$ in type IIA theory
(if duality is 
performed along one of 4 `unboosted' directions), which  
 is the type IIA image of  boosted 11-dimensional 5-brane 
reduced along an isometric coordinate orthogonal to it,\foot{We shall use 
subscript to 
 indicate the existence of a number of periodic isometric
directions  orthogonal to p-brane,  on which harmonic functions in the
solution do not depend.} 
$(5 + \upa)_1$; (b) bound state of 5-brane and fundamental
string  $5 + 1$   (if duality is performed along the boosted direction), 
which is a  reduction of the $D=11$
$5 \bot 2$ solution  describing an orthogonal intersection 
of 5-brane and 2-brane over a  string   \tset.

(2) R-R 5-brane of IIB theory 
with a momentum flow along one of its directions,  $5_{R} + \upa$, 
is $T$-dual to: (a) $(4 + \upa)_1$  which is the reduction of $(5 +
\upa)_1$
in $D=11$; (b) $4 \bot 1$  which  is  
another   reduction of $ 5\bot2$ in $D=11$. 

Let us first ignore the momentum flow and consider 
the $(q_1,q_2)$ bound state $5_{NS}+5_{R}$ obtained by applying the 
$SL(2,R)$ rotation to $5_{NS}$ of \refs{\duflu,\chs}. 
 Its $T$-dual
 can be interpreted as a   1/2
supersymmetric non-threshold 
 bound state 
of the solitonic 5-brane and R-R 4-brane in type IIA theory, 
i.e.  $5+ 4$. 
The image of the latter  in $D=11$   is $5_1$, i.e.
the  5-brane in $M^{11}= S^1_{y_2} \times M^{10}$, 
 reduced down to $D=10$  along the 
mixture of the isometric 
direction  $y_2$ transverse to the 5-brane, and a  
5-brane direction $y_1$ (which  will be 
orthogonal to the 4-brane within the type IIA 5-brane).
These two directions form the 2-torus  related to type IIB
$SL(2,Z)$ duality
 which is the analogue 
of $T^2$  in  the previous section.

The  $D=11$   $5_1$  background   can  be 
 viewed  as  the  limiting case of the $5\bot2$ solution
  when  the  charge of
 2-brane is set equal to zero (with $y_2$ being the 2-brane direction 
 orthogonal to  5-brane).  The reduction is done 
 along the $(q_1,q_2)$ cycle of the 2-torus around which 
  the 2-brane is wrapped. 
 When $q_1=0, q_2=1$
 we get simply $4_1$  (the reduction goes along 
 a  5-brane direction $y_1$  common to $5$ and $2$).
 When   $q_1=1, q_2=0$ we get the type IIA  5-brane 
 (the reduction goes along the 2-brane direction $y_2$ 
 transverse to the 5-brane).

Next, we can add momentum along some special direction on 5-branes,
 i.e.  start with type IIB background  $5_{NS}+5_{R} + \upa$.
  It is  natural  to add  also  
  a string  lying along the same direction, 
  i.e. to consider the $SL(2,Z)$ rotation
  of the 1/8 supersymmetric  background  $ (1+5)_{NS} + \upa$  \TT,
   representing 
  the threshold bound state of a  boosted fundamental 
  string and  solitonic 5-brane.
  The resulting  $(q_1,q_2)$ configuration 
  $ (1+5)_{NS} + (1+5)_{R}+ \upa$ 
  ($n=3,4,5,6$; $y_2$ is reserved for the eleventh coordinate) 
 $$ 
ds^2_{10 B  } =  H_5^{1/2} H_{15}^{1/2}
 \big( H_1\inv H_5\inv  \big[  -dt^2 + dy^2_1 + V (dt-dy_1)^2\big]
  + H_5\inv dy_n dy_n    + dx_i dx_i \big)\ , 
$$ 
\eqn\onef{e^{2\p}= H_1\inv H_5\inv H_{15}^2  \ , 
\ \ \ \  \ \   \chi =  \sin \t  
 \cos \t\  (H_1-H_5) H_{15}\inv    \ ,  \ \ \ \ V =  {\Q\ov r^2} \ , 
 }
   $$
 dB^{(1)} + i dB^{(2)} =  (\cos \t + i \sin\t ) (dH_1^{-1} \wedge dt\wedge dy_1
 + *d H_5 ) \ ,   $$
 $$ H_{15} \equiv  {\rm cos^2}\t \ H_5 + {\rm sin^2}\t \ H_1  \ , 
 \ \ \  H_{1,5} = 1 + {Q_{q (1,5)} \ov r^2} \ ,  \ 
 \  \ \ Q_{q(1,5)}  = Q_{1,5} \sqrt 
 {q_1^2 + q_2^2}  \ , $$
 is  parametrized by 5 charges 
  and includes as special cases
  $ 1_{NS} + 1_{R}+ \upa$ (i.e. reduces to 
  \johhn\ when $H_5=1$, $K=H_1, \ \K= H_{15}$, \ $V=0$)
  and $ 5_{NS} + 5_{R}+ \upa$ ($H_1=1$).
  Its  mass is $M = {2\pi^2\ov \k_5^2} 
  \sqrt { q_1^2 + q^2_2} ( Q_1 +Q_5)$. 
  $T$-duality  along $y_1$ converts this into a complicated 
   type IIA solution which can be interpreted
  as a 1/8 supersymmetric non-threshold bound state of
 a 5-brane, 4-brane,  string and a boosted 0-brane, i.e.
 $5+4+1+0\ma$.

  Lifting the latter to $D=11$  along $y_{11}\equiv y_2 $ 
  and setting first the momentum to zero (i.e. $V=0$)
   we find  the following metric
   $$  
   ds^2_{11}= H^{2/3}_5 \big(- H_1\inv H_5\inv dt^2
   + H_1  H_{15}\inv  \left[ dy_1 -   \cos \t \  (H_1-1) H_1^{-1}  
   dt\right]^2  $$
   $$  +\  H_5  H_{15} \left[ dy_{11} 
     -  \sin \t\ (H_1-1) H_{15}^{-1} dt
         + \sin \t  \cos \t\  (H_1-H_5) H_{15}^{-1} dy_1 \right]^2  
  $$ \eqn\moo{
   +\  H_5\inv dy_n dy_n    + dx_i dx_i \big)\ ,   }
  which reduces to \memoo\ when there are no 5-branes $H_5=1, H_1=K$. 
  This can be  re-written  simply as 
  the  `rotated' $5_1$-brane 
  \eqn\fff{
  d s^2_{11} =  H^{2/3}_5 
 \left[  H_5\inv   (-  dt^2   + dy_1'^2  + dy_n dy_n) +
 dy'^2_2  + d x_i dx_i \right] \ , }
 where $y_1',y_2'$ are the rotated coordinates  \rott.  
  As expected, the same turns out to be true for non-vanishing momentum:
  the background one finds  can be put into the form (cf. \memo;
  $y_2\equiv y_{11}$) 
 $$ 
d s^2_{11} =  H^{2/3}_5 H^{1/3}_2
 \big(  H_5\inv H_2\inv   [-  dt^2  
 +  dy'^2_1    + 
   W (dt -  dy'_1)^2 ] $$ \eqn\vv{ 
    + \  H_5\inv  dy_n dy_n +    H_2\inv  dy'^2_2       
+  dx_i dx_i \big) \ ,    } 
$$ dC_3 
 =   dH_2\inv\wedge  dt \wedge dy_1 \wedge dy_2  
 + *dH_5  \wedge dy'_2 \ ,  \ \ \ \   H_2\equiv 1 + V \ ,  \ \ 
 W\equiv H_1-1 \ .  $$ 
This  is just  a    `rotated'   version  of $5 \bot 2 + \upa$, i.e.
of the intersection of the 2-brane 
and 5-brane boosted along the common string 
\refs{\tset,\klts} where the momentum now flows along the $(q_1,q_2)$
cycle of the $(y_1,y_2)$-torus  around which  
  the 2-brane is wrapped. 
  
Special cases of the corresponding  type IIA and type IIB solutions 
include  (depending on direction of reduction): 
  $2\bot 5 \la 2\bot 4 $ in IIA, related by $T$-duality to 
   $ (1 + 5)_R$ in IIB;  
  $2\bot 5 \la 1\bot 4 $ in IIA, related by $T$-duality to 
   $  5_R + \upa $ in IIB; 
$2\bot 5 \la 1 + 5  $ in IIA, related by $T$-duality to
 $  5_{NS}  + \upa $  or 
$ (1 + 5)_{NS}$ in IIB.
Reduction down to $D=5$ gives a family of 
 regular extremal 
 black hole
 solutions related to the simplest  NS-NS \TT\ and R-R \CM\ ones
 by  U-duality. Reduction along the four 5-brane directions
 gives a family of solitonic strings in $D=6$.

We conclude that  the  $5\bot 2 + \nea$  background \vv\ 
provides  a unified  $D=11$ description of 
 various  1-brane and 5-brane bound state configurations
of type IIB theory.\foot{In addition to the special cases of 
$(1+5)_{NS} + (1+5)_{R}+ \upa$  there are also 
1/2 supersymmetric non-threshold bound state 
of the type $1_{NS} + 5_R$, i.e. the fundamental string lying on R-R 5-brane
(and its S-dual $ 1_{R} + 5_{NS}$).   Such solution can be constructed by
$T$-duality in 4 transverse directions from  $(1_{NS} + 1_R)_4$.
Its $T$-dual (along the string direction) 
counterpart  in IIA theory  is  $4$-brane finitely  boosted 
in transverse dimension, or $4\ma$.}

\subsec{ \bf  
$2+0$ and  $5+0$ bound states  as  transversely boosted M-branes}

The  type IIA 
1/4 supersymmetric  threshold bound states of a string and 0-brane, 
$1 + 0$,  and of 4-brane and 0-brane,  $4+0$,  
 have  a simple  origin in $D=11$: the corresponding solutions are
 readily   obtained 
 by  dimensional
reduction  from 
extremal M-brane configurations  with momentum along the brane 
$2 + \up$ and $5 + \up$, or 
black  M-branes 
infinitely boosted in  one direction  {\it parallel} to the brane 
\refs{\tset,\tser}.\foot{For a discussion of  alternative  $D=11$  embeddings  of $1 +0$ and $4+0$ and  also 
the embedding of $8+0$ see \paptkk.}

The existence of $1_{NS} + 1_R$ type IIB bound state  suggests 
by $T$ and $SL(2,Z)$  duality the existence of  other 1/2 supersymmetric
{\it non-threshold}  type IIA bound states that can be 
interpreted as $2+0$, 
$5+0$, etc.\foot{The existence
 of related  D-brane  bound states was  noted in \refs{\jch}. 
 For a discussion of $2+0$ bound state 
in D-brane representation see \lifsh.
The corresponding classical   solutions cannot be obtained by 
formal $T$-duality transformations \beh\ of the standard R-R p-brane
backgrounds.}
In view of the above   identification 
of the $D=11$  counterpart  of the string-string 
bound state  background it is of interest (in particular, 
in connection with \banks)  to  enquire about the 
$D=11$ origin  of such bound states. 
As we shall demonstrate below, the corresponding classical solutions 
are indeed very simple: they are just a  coordinate  (Lorentz) 
transformation 
of the  extremal M-brane  background and 
 describe  an  M-brane  {\it finitely} boosted in the  direction
 {\it transverse} to its internal space.\foot{The  transversely  boosted 
 extremal p-branes  are   physically different from the 
 static ones as the Lorentz 
 transformation  is globally non-trivial,
   just as in the case of the 
 longitudinally  boosted non-extremal p-branes
   \horow\ (extremal p-branes are, of
 course, invariant under longitudinal boosts).}
  Dimensional reduction 
 along this  direction 
 gives $2+0$ and $5+0$ type IIA solutions.  These solutions 
 interpolate between the standard 
 static  M-brane (zero boost) and plane wave (infinite boost). 
  As will follow from the relation to the
  $1_{NS} + 1_R$ solution, the  coefficient in the 
 harmonic function 
  of  the transversely boosted brane 
 depends on the boost parameter and goes to zero 
 in the limit of the infinite boost when 
 the $D=11$ solution becomes just a plane wave (and the corresponding
 type IIA solution becomes a 0-brane).

We shall start with the explicit construction of $2+0$ solution
 by  duality transformations   from $1_{NS} + 1_R$  background 
 of \john\ and then lift the result  to $D=11$.
Assuming that the harmonic function $W$ in \johhn\ does not 
depend on two of the transverse coordinates (to be denoted as $y_2$,
$y_3$), i.e. starting with \johhn\ with two extra isometries,
 $(1_{NS} + 1_R)_2$, we may apply the 
$T$-duality in these  isometric directions to obtain a
 type IIB solution which has  an  interpretation 
 of a bound state of a fundamental string and a 3-brane, 
$1_{NS} + 3$,  
\eqn\hhn{
ds^2_{10 B} =   \K^{1/2} [ K\inv ( -dt^2 + dy_1^2) + 
\K\inv (dy_2^2 + dy_3^2) +  dx_s dx_s ]  \ , }
$$ e^{2\p }=   K\inv \K   \ , \ \ \ \   \chi =  0 \ , \ \  \ \ 
D_{ty_1y_2y_3} = \sin \t\ W \K\inv \ , $$
$$ 
 B^{(1)}_{t y_1 }  =  - \cos \t\ W  K^{-1}  \ , \ \  
  B^{(2)}_{y_2 y_3 }  =  \sin \t  \cos \t\  W \K^{-1}  \ ,   $$
  where $K$ and $\K$ have the same form as in \defo, and we do not
   list other nonvanishing 
   components of the 4-tensor determined by  self-duality of 
   its  field strength. 
  This background interpolates between the fundamental string 
 $1_2$ ($q_1=1,q_2=0$, i.e. $\t=0,\ \K=1$) and  the 3-brane
  ($q_1=0,q_2=1$, i.e. $\t= {\pi \ov 2},\ K=\K$). 
The simple  $SL(2,Z)$ transformation  
 which  interchanges  the NS-NS  and  R-R  strings 
 and leaves 3-brane invariant then gives  similar 
   $1_R +3$ background
\eqn\hhen{
ds^2_{10 B} =   K^{1/2} [ K\inv ( -dt^2 + dy_1^2) + 
\K\inv (dy_2^2 + dy_3^2) +  dx_s dx_s ]  \ , \ \  e^{2\p }=   K \K\inv 
  \ , }
  with the same values of $D_4, \chi$ and interchanged values 
  of $B^{(1)}$  and $B^{(2)}$.
The final  step is to apply $T$-duality in $y_1$ direction
\eqn\hen{
ds^2_{10 A} =   K^{1/2} [ - K\inv dt^2 + 
\K\inv (dy_2^2 + dy_3^2)   + dy_1^2  +  dx_s dx_s ]  \ , }
$$ e^{2\p }=   K^{3/2}  \K\inv   \ , \ \  \ \ \ 
C_{ty_2y_3} = - \sin \t \ W \K\inv \ , $$
$$ A_{t }  =  - \cos \t\ W K\inv \ , \ \  \ \ \ 
  B_{y_2 y_3 }  =  \sin \t\  \cos \t\  W \K^{-1}  \ .   $$
This  background  can be interpreted as $(2+0)_1$
since it interpolates between the 0-brane ($q_2=0,\  \K=1$) 
and  the 2-brane wrapped around $y_2,y_3$ 
 ($q_1=0, \ K=\K$) in the space with one extra isometry
($y_1$). As $y_1$ does not play any special 
role,  the restriction of  the  extra isometry can be relaxed
(i.e. $y_1$ can be  added to  $x_s$)
obtaining the $2+0$ background  which is spherically symmetric in all 
7 transverse coordinates. 

This  type IIA solution can now be lifted to $D=11$
(cf. \memo,\memoo)
$$
ds^2_{11} =   \K^{1/3} [ - K\inv dt^2 +  K\K\inv (dy_{11}- 
\cos \t\ W K\inv dt)^2     + 
\K\inv (dy_2^2 + dy_3^2)   +  dx_i dx_i ]  \ , 
$$
\eqn\hene{ C_3 = - \sin \t\ W \K\inv   dt\wedge dy_2 \wedge dy_3 
   +         
 \sin \t \cos \t\ W \K\inv   dy_{11} \wedge dy_2 \wedge dy_3 \ . 
}
 This can be re-written  simply as 
 \eqn\hnn{
ds^2_{11} =  \H_{2}^{1/3} \big[  \H_{2}\inv  (-  d\td t^2 + 
dy_2^2 + dy_3^2) 
  +  d\td y^2_{11}  +  dx_i dx_i \big]  \ , }
$$ dC_3 =   d\H_{2}\inv \wedge   d\td t\wedge dy_2 \wedge dy_3  \ ,   \ \
\ 
\ \ \ 
\H_{2} \equiv \K = 1 + {\rm sin}^2\t\ {\Q_q\ov r^6} \ , 
 $$
 where, as in \heq, 
$$ \td t\equiv {1\ov \sin \t} ( t - \cos \t\ y_{11}) \ , \ \ \ \ 
\td y_{11} \equiv  {1\ov \sin \t} ( y_{11} - \cos \t\  t ) \ ,
\ \ \ \   -\td t^2 + \td y^2_{11}  = - t^2 + y^2_{11} \ . 
$$
 This  is just the  2-brane  \mem\  with
 harmonic function 
  $ H_2 = \H_{2}  $  
 boosted  to a subluminal  velocity  $v= \cos \t \leq  1 $ 
 in the isometric  transverse  direction  $y_{11}$ 
 (this background  will be denoted as
  $2\ma$). 
 This matches perfectly with the $2+0$ interpretation of the
 corresponding type IIA solution!
 
The background \hnn\ interpolates between the standard extremal 2-brane
and a plane wave:   for  $q_1=1,q_2=0$ (i.e.  infinite boost $v=1, \ 
 \sin\t =0,  \ 
\H_{2}=\K=1$)
we get (by carefully taking the limit $\sin\t \to 0$)
 the  gravitational wave along $y_{11}$ direction which 
reduces to 0-brane in $D=10$,  while  for 
 $q_1=0,q_2=1$ (i.e. zero boost  $v=0, \ \sin \t = 1, \ H_{2}=\K=
  1 +  {Q\ov r^6} , \ Q\equiv \Q) $
we find  just the static $2_1$ M-brane  which reduces  to type IIA 2-brane.  
For general $v$ \hnn\ represents a 2-brane with transverse momentum 
(in the $y_{11}$ direction).
 The energy is  given by the usual 
relativistic expression
$E^2 = M^2 + p^2   \sim  {\rm sin}^2 \t \ \Q^2_q  
+  {\rm cos}^2 \t\ \Q^2_q=
 \Q^2_q$, with the first (second) term being zero in the  infinite (zero)
 boost limit.
 This gives a `kinematic' interpretation to the mass formula 
 of a non-threshold BPS bound
 state.\foot{This  is also  related to the 
 explanation 
  of  the tension of the bound state of IIB strings
  in terms of a
 D-string  (Born-Infeld) action 
 \refs{\schmid}.}

Let us note also that there exists a more direct way of relating the 
static  2-brane  in the  $D=11$ space 
with an isometry along $y_{11}$  (i.e. $2_1$) and the type IIB string-string
solution  \johhn: one is to reduce down to $D=10$ along the `rotated'
 direction of the  $(y_3,y_{11})$ torus, i.e. to  wrap the 2-brane around
 its generic cycle,   
 obtaining the $2+1$ non-threshold bound state
 of type IIA theory,  and then to  make $T$-duality along $y_3$. 
 The resulting  IIB background is $(1_{NS} + 1_R)_1$, i.e.
 the string-string  bound state  in the 
 space  with an isometry along the $y_3$ direction 
 orthogonal  to the string direction $y_2$.

The above  discussion admits  the following   generalization.
If we start with the {\it boosted} version \jhn\ of the type 
IIB $(q_1,q_2)$ string  and apply duality transformations 
  we find the following sequence
of 1/4 supersymmetric non-threshold bound states:
$$ (1_{NS} + 1_R + \up)_2\  \la \ 1_{NS} +\up +  3\  \la 
\ 1_{R} +\up +  3\ \la \ 2 \bot  1  + 0  \ . $$
The last one is  a   generalization of the 1/4 supersymmetric
threshold  configuration $2\bot 1$ 
of $2$-brane orthogonally  intersected by  a  fundamental string
 over a  point  where an extra 0-brane is now  placed. 
 Recalling that  $2\bot 1$ is the dimensional reduction
  of $2\bot 2$ in $D=11$
 \refs{\papd, \tset}   it is clear that $2 \bot  1  + 0$
 should be a dimensional reduction of  $2\bot 2 \ma$,  
 where one 2-brane is finitely  boosted   in the 
 transverse direction which is  a  longitudinal direction
 of the other 2-brane, i.e. 
$$
d s^2_{11} =  
\H^{1/3 }_{2(1)}  H^{1/3}_{2 (2)}  
\big[ - \H^{-1}_{2(1)}  H^{-1}_{2(2)}  d\td t^2  
+  \H^{-1}_{2 (1)} (dy_2^2 + dy_3^2) 
  + H^{-1}_{2 (2)} (d\td y_{11}^2 + dy_1^2) 
  +  dx_s dx_s \big]  ,  $$  \eqn\twot{  
   C_3 =   \H^{-1}_{2(1)}  d\td t\wedge dy_2 \wedge dy_3 
  +  H^{-1}_{2(2)}  d\td t\wedge d\td y_{11} \wedge dy_{1}  \ , }
  where  $\td t, \td  y_{11}$ and $\H_{2(1)}$
  are defined in a similar way as above.
  $ y_2,y_3$  and $y_1,y_{11}$ are the coordinates 
  of the two 2-branes, and the first  brane  is boosted along 
  $y_{11}$ direction of the second.
  In the infinite boost limit ($\sin \theta =0 , \ 
  \H_{2(1)} =1$)  this becomes $2+ \up$  which  reduces down to
  $1+0$ in $D=10$.

Another chain of duality  transformations 
$$(1_{NS} + 1_R)_4\ \la\  1_{NS} + 5_R\ \la\ 1_R + 5_{NS}\
 \la \ (0+ 5)_1  \la \ 5+0 \ $$    
leads to the type IIA  bound state of a solitonic 5-brane and a 0-brane.
Its image in $D=11$ is  simply  a  5-brane finitely  boosted in the 
transverse isometric direction, 
i.e.  $5 \ma$. The corresponding background is
the obvious analogue of \hnn.  


 Starting with  $(1_{NS} + 1_R + \up)_4$
we arrive at 1/4 supersymmetric type IIA solution $ 5 + 1 +0$ 
describing the  bound state  of a 5-brane with a fundamental 
string  and a  0-brane. This is a non-threshold bound
state, except for   the  special cases $1+0$ and $5 +1$ which are 
 at threshold. Its $D=11$ image  is  $2\bot 5 \ma$,
i.e. the orthogonal intersection of 5-brane and 2-brane over a string
with 5-brane  finitely  boosted  along the direction of 2-brane orthogonal
to it. In the limit of the infinite boost   the 5-brane 
charge goes to zero  and this 
 configuration becomes just the longitudinally boosted 
 2-brane,  $2+ \up$  (which  reduces to $1+0$). 
The corresponding solution is the  direct   analogue  of the above
expressions (cf. \vv) 
$$
d s^2_{11} =  \H^{2/3}_{5}  H^{1/3}_2
 \big[  \H_{5}\inv H_2\inv   (-  d\td t^2  
 +  dy^2_1 )    +  \H_{5}\inv  dy_n dy_n +    H_2\inv  d\td y^2_{11}        
+  dx_i dx_i \big] \ ,    $$
\eqn\rfe{ dC_3 
 =  dH_2\inv  \wedge d t\wedge  dy_1 \wedge d\td y_{11} 
 + *d\H_5  \wedge d\td y_{11} \ ,  \ \ \ \ 
 \H_{5} \equiv 1 + {\rm sin}^2\t\ {Q_{q5}\ov r^6} \ .  }
where the 2-brane coordinates are $y_1$ and $y_{11}$.

Although the   $D=10$ solutions  $1_{NS} + 1_R$, $2+0$  and $5+0$ 
(or  their generalizations -- $1_{NS} + 1_R + \up $, $2\bot 1+0$ 
 and $5+1+0$) 
are related by  $T$ and $SL(2,Z)$  dualities, 
 their $D=11$ counterparts  --
  the wave   along generic cycle of 2-torus,  $2 \ma$ and $5 \ma$ 
  (or 2-brane boosted in $(q_1,q_2)$  longitudinal
  direction $2 + \nea$,  $2\bot 2 \ma$ and $2\bot 5 \ma$) 
  have  different structure.
  
  There is one more example of this dichotomy:
  $2+0$ is $T$-dual to   $4+2$ which  in turn results  upon  
  dimensional reduction \refs{\grepap}  from   
another $D=11$   solution which   
interpolates between the 2-brane
and 5-brane  backgrounds. It    can be interpreted as
a  2-brane lying within a  5-brane    \iz\ (see also 
\refs{\papadop,\grepap}).  In the notation  used here it is given
by (cf. \vv,\hene)\foot{This form of the  $2+5$  solution  
 is found  if one  first obtains  $4+2$  by   dualities 
from our basic starting point  $1_{NS}+1_R$ \johhn\ 
and then lifts  $4+2$  to $D=11$.}
$$d s^2_{11} =  K^{1/3} \td K^{1/3}
 \big[  K\inv    (-  dt^2  + dy_1^2 + dy_2^2) 
 +  \td K\inv    (dy^2_3 + dy^2_4 + dy^2_5)  +  dx_i dx_i \big] \ ,   
 $$ 
 \eqn\izq{K= 1 + W \ , \ \ \ \    \td K = 1 + {\rm sin}^2\t\  W \ , \ \ \ 
 W= {Q_{q} \ov r^3} \ , } 
 $$ dC_3 
 =  \cos \t\ dK\inv  \wedge dt\wedge  dy_1 \wedge dy_2 
  + \sin \t\  *d K   -   
  6 \cot \t \   d \K\inv  \wedge dy_3 \wedge dy_4  \wedge dy_5 \ . $$ 
Since for    $\t=0$  this background  reduces to 
  the  2-brane ($\K=1, K=H_2$) 
  and for $\t=\pi/2$  to the 5-brane ($K=\K=H_5$), 
   and in view of  the close analogy with 
   similar solutions discussed above, 
   it  can be  interpreted  
as  describing  their 
1/2 supersymmetric  non-threshold  bound state $2+5$.
Reducing to $D=10$ along  a 5-brane coordinate  one finds the
 $4+2$ solution, while the 
reduction along a 2-brane coordinate gives a similar 
bound state of a 4-brane and a 
fundamental string $4+1$. The latter, $(4+1)_1$, 
  is $T$-dual to  a non-threshold type IIB bound state of a R-R 5-brane and
  a fundamental string,  $5_R + 1_{NS},$
  which thus has $5+2$ bound state as its $D=11$ counterpart. 

It remains an interesting question  whether  the above $D=11$ 
 solutions with boosts are somehow connected  to this static solution
 in an intrinsically  11-dimensional way
 (which goes beyond the  connection by string dualities 
 of dimensionally reduced backgrounds  with KK modes dropped 
 out).
 This  may give a hint about  a  relation between  M-branes  and
 waves,  similar to the $T$-duality relation  
 between fundamental strings and
 waves.\foot{Just like $D=10$ solutions with at least one isometry 
  are related by $T$-duality, $D=11$ solutions with at least 
  two isometries are related by similar symmetry transformations
  \berg\  of 
  supergravity actions. In string theory,  $T$-duality 
  is not only a symmetry of the  leading term in the  effective action but 
  is present to  all orders in string perturbation theory 
  as  it has  its origin in 2-d duality of world-sheet action.
  The question is about analogous exact symmetry of M-theory.}

The $D=11$  solutions discussed above  have many generalizations.
One may combine  the threshold bound state 
intersecting solutions  with
longitudinal momentum along common string direction  
\refs{\papd\tset\klts\gaunt -\pope} 
with  boosts of branes in transverse directions. 
One  may also combine intersection
 solutions with static non-threshold $2+5$
bound states 
of \iz\  (some of such solutions were discussed in \costa).\foot{We
expect that one of such solutions reduces to $4+ 2\bot 2+0$ 
bound state discussed in D-brane approach in \lifsh.}

\newsec{Correspondence between  microscopic BPS states of 
type IIB strings and quantum  supermembrane}

\def\n { {(k,m)} }
\def\mn { {(-k,-m)} }

\def\ww {\omega _{km } }

\def\rs { ${\bf R}^{10}\times S^1 $}
\def\rss { ${\bf R}^{9}\times S^1 \times S^1 $}

It was suggested  in  \refs{\john,\schwa}
that the spectrum 
of   BPS states of type IIB  $(q_1,q_2)$ string on a circle 
 should 
be in correspondence with 
  the BPS spectrum of the fundamental supermembrane wrapped around 
a  2-torus with a momentum along the $(q_1,q_2)$ cycle of the 
torus. What was shown in  \refs{\john} is that  
the zero-mode parts of the spectra  do match.  Since the states in
question are, in general, oscillating states  
it is important to check  that the oscillator parts in 
the masses  (and the constraints) also agree. 
 
 As was argued in \john, 
the BPS spectrum of $(q_1,q_2)$ string should be the same 
as that of perturbative $(1,0)$ string  provided  the tension
is  rescaled by $\sqrt{q_1^2 + e^{-2\p_0} q_2^2}$. 
The  problem  which  we  address below 
is  how to determine  the {\it oscillator }
part of the quantum  supermembrane  BPS spectrum, given that this 
is a complicated   interacting  theory (assuming it is well defined
as a quantum theory in the first place).

For a membrane wrapped around  the torus, the light-cone gauge 
  Hamiltonian $H$ turns out to contain
a `free' (quadratic) $H_0$ part  and an  interaction 
 part $H_{\rm int}$. The theory can be solved in a special limit of large torus
 area in which the interaction  term $H_{\rm int}$ drops out.
 Our main assumption will be  that the masses of  the BPS states of
 wrapped supermembrane do not receive corrections in
decreasing the torus area while keeping fixed the torus modular 
parameters and the $D=10$ string tension  $T_2 = 1/2\pi\a'$
 (in this supersymmetric problem one expects that 
  masses of  BPS states with  given 
   charges should not receive quantum corrections depending 
   on continuous parameters like radii of the 2-torus, i.e.
   masses should remain the same as  for  $H_{\rm int}=0$). 
Solving the gaussian theory with $H=H_0$  we find
 that  the resulting oscillating membrane BPS  states 
 do, indeed, have the same  masses as the BPS states
 of type IIB strings. This   provides   strong support for   
  the suggestion  of  \john.


It should be emphasized that this  correspondence  between the BPS 
 states of $(q_1,q_2)$ string and wrapped membrane  with `quadratic' 
 Hamiltonian  is quite non-trivial, 
 as the `quadratic' approximation  reproduces  not only
  the perturbative 
 NS-NS  $(1,0)$ part of string spectrum but also the  R-R  $(0,1)$ 
 and the  general $(q_1,q_2)$  parts as well.
 It is remarkable that these (non-perturbative) bound states
 of type IIB superstring theory admit an interpretation 
  in terms
 of excitations of a fundamental supermembrane with certain
 values of Kaluza-Klein momentum and winding number.

This 
complements  the  relation  between 
the $(q_1,q_2)$ string and $D=11$  membrane with momentum flow 
along the $(q_1,q_2)$ cycle of 2-torus 
established  at the macroscopic effective field theory level 
in Section 2.  In the case of the fundamental type II strings 
the classical `string+wave'  solutions 
are in   correspondence 
with   microscopic BPS states \refs{\dgh,\calmalpeet,\dabwal},  
with momentum being carried  by chiral (right-moving)
 string oscillations. 
This  suggests that the `membrane+wave'  solution \memo\ 
(more precisely, its oscillating generalization mentioned in Section 2)
 should also correspond to microscopic membrane BPS states with
 longitudinal momentum  carried by  membrane
oscillations 
 propagating along {\it one}  direction -- the $(q_1,q_2)$ cycle of the torus. 
 Indeed, as shown below, these are the  membrane states whose masses 
 coincide with the masses of the BPS states of $(q_1,q_2)$ string.

\subsec {\bf Supermembrane theory on \rss }
The membrane states investigated here are excitations of a membrane
with non-trivial winding number around the target-space  torus.  
The spectrum of the light-cone membrane Hamiltonian
in this case is discrete (an earlier study of the spectrum on \rss\ is 
in ref.~\dufi).


Let $X^{1}$ and $X^{2}$ represent the compact coordinates, with
periods $2\pi R_{1} $ and $2\pi R_{2}.$
\  $X^{2}$  will  be viewed as an  extra  `eleventh' coordinate
 absent in  superstring theory. 
Let  $\s , \rho \in [0,2\pi )$ be the world-volume spatial
 dimensions.
The transverse, single-valued coordinates $X^i$, $i=3,..., 10$,
and the canonical momenta $P^i$ 
 can be expanded in a complete set of functions on the 
 torus\foot{We shall  use  the indices $k, l$  for the 
  Fourier components
in $\s $, and indices $m,n,p $ for Fourier components in $\rho $.}
\eqn\adim{
X^i (\s , \rho )=\sqrt{\a'  } \sum _{k,m} X_{(k,m)}^i e^{i k \s+im\rho  }\ ,\ \ \ 
 P^i(\s, \rho )={1\ov (2\pi)^2 \sqrt{\a' } } \sum _{k,m}  
P^i_{(k,m)} e^{ ik \s + im\rho}\ , 
}
where 
\eqn\aadd{
\a' \equiv ( 4\pi ^2 R_{2} T_3 )^{-1}\ , \ \ \ \ \ \ \  [T_3]= cm^{-3}  \ . 
}
The canonical commutation relations imply 
\eqn\ccnn{
[X_{(k,m)}^i, P^j_{(k',m')}]=i\delta _{k+k'} \delta _{m+m' }\delta ^{ij}\ .
}
For a  membrane  wrapped
around the rectangular  target-space  torus  so that 
\eqn\aatt{
X^{1}(\s +2\pi ,\rho) = X^{1}(\s ,\rho ) + 2\pi w_{1} R_{1} \ ,
\ \ \ \  X^{2}(\s ,\rho +2\pi ) = X^{2}(\s ,\rho ) + 2\pi w_{2} R_{2} \ , 
}
one has 
\eqn\uno{
X^{1}(\s,\rho )=w_{1}R_{1}\s +\tilde X^{1}(\s ,\rho) \ ,\ \ \ 
\tilde X^{1}(\s, \rho )=  \sqrt{\a'  } \sum _{k,m} X_\n^{1} e^{ik\s +im\rho }\ ,
}
\eqn\dos{
X^{2}(\s,\rho )= w_{2}R_{2}\rho +\tilde X^{2} (\s,\rho )\ ,
\ \ \ 
\tilde X^{2} (\s,\rho )=\sqrt{\a'  }\sum _{k,m} X_\n^{2} e^{ik\s +im\rho }\ .
}
The winding number that counts how many times the toroidal membrane
is wrapped around the target-space torus is
\eqn\wiwi{
w_0={1\ov 4\pi^2 R_{1} R_{2} }\int d\s d\rho \ \{ X^{1}, X^{2}\} 
= w_{1} w_{2} \ , \ \ \ \ \ \{ X, Y \} \equiv \partial _ {\s } X  
\partial _ {\rho } Y - \partial _ {\rho } X  
\partial _ {\s } Y 
\ .  }
A membrane with $w_0\neq 0$ is topologically protected against
usual supermembrane instabilities  \dewit.
It is convenient to  choose  the light-cone gauge to remove the single-valued
part $ \tilde X^{1}$ of $X^{1}$,   i.e.  $X^+=x^+ +\a' p^+ \tau $,\ 
$X^\pm \equiv (X^0 \pm \tilde X^{1})/ \sqrt {2} \ .$
The standard light-cone Hamiltonian of the supermembrane is given by
\refs {\bst , \dewitt } 
$$
H=H_{\rm B} + H_{\rm F} \ ,
$$
\eqn\lcg{
H_{\rm B} =  
2\pi ^2 \int d  \sigma d\rho \left [
 P _ a   ^ 2 + \ha 
{ T_3^2 }
( \{ X ^ a, X ^ b \} ) ^ 2\right ] \ ,
}
\eqn\lcgh{
H_{\rm F} =- T_3p^+  \int d  \sigma d\rho \ 
 \bar \theta \Gamma _ a \{ X ^ a  , \theta \} \ ,
} 
where  $a=1,2,...,10$ and  $\td X^{1}= 0 $.
Here $\theta _\a $ $(\a =1,...,16 )$  is a  real $SO(9)$ spinor.
The (mass)$^2$  operator $M^2=2p^+p^--p_{a}^2$ is given by 
$2 H - p_{a}^2 $, where $p_a$  is the 
center-of-mass momentum of the membrane, $p_a=\int d\s d\rho P_a $.
For simplicity of presentation, in what follows we will omit the
 fermionic terms  in  the formulae  
(their inclusion  is straightforward, see,  e.g.,  \refs {\dufi, \russo }).

The Hamiltonian has a residual symmetry associated with area-preserving
 diffeomorphisms, which can be fixed by setting
 \eqn\aasd{
 \td X_{2} (\s ,\rho )= X_{2}^C(\rho )\equiv \sqrt{\a'  }
 \sum _m X_{2 (0,m)} e^{im\rho }\ . 
 }
One can split  $P_{2} (\s ,\rho )= P_{2}^C + \hat P_{2}$, where
$P_{2}^C=P_{2}^C (\rho )$  belongs to the (Cartan) subspace
generated by $e^{im\rho }$, and $\hat P_{2}$ to  the complement.
The local constraints can be solved for $\hat P_{2}$ in terms of 
$\td X_{2} $ and the transverse coordinates and momenta $X_i, P_i  $.  $\hat P_{2}$  can be ignored in 
  the problem of BPS masses  we are interested here 
(in the large radius limit 
considered below  $\hat P_{2}$   gives
subleading contributions of order $O({1\ov R_2})$, see ref.~\russo ).

Separating the 
winding contributions, 
 we  can put   the  Hamiltonian  in the form 
\eqn\hhh{
H_{\rm B}=H_0+H_{\rm int}\ ,
}
\eqn\hhcc{
H_0= 2\pi ^2 T_3^2 A w_0^2 + 2\pi ^2 \int d\s d\rho
\big[ P_a ^2 + T_3^2 R_{2}^2 w_{2}^2\big( \del_\s X_a\big)^2 +
T_3^2 R_{1}^2 w_{1}^2 \big( \del_\rho X_a \big)^2\big]\ ,
}
\eqn\hhuu{
H_{\rm int}= \pi ^2 T_3^2\int d\s d\rho 
\bigg[ 4 w_2 R_2 \del_\rho X_2 \big( \del_\s X_i\big)^2 +
\big(\{ X^a, X^b\} \big)^2 \bigg]\ ,
\ \ \ \ \ \ A=4\pi ^2 R_{1}^2 R_{2}^2\ ,
}
where $i=3,...,10$ and $a,b=2,...,10$, and 
$X_{2}= X_{2}^C$.
As discussed in \russo , the spectrum of this Hamiltonian is discrete provided
$w_0=w_1w_2 \neq 0 $ (for $w_0=0 $
one or both  terms in the quadratic part of the potential vanish,
flat directions remain, leading to a continuous spectrum).
Inserting the expansions \adim , \uno\ and \dos , we obtain
\eqn\enmo{
\a'  H_0={1\ov 2} \sum _{k,m} \big[ P_\n^a P^a_\mn
+\ww ^2 X^a_\n X^a_\mn \big]
+ {R_{1}^2 w_0^2\ov 2\a'  }\ ,
}
\eqn\nome{
\a'  H_{\rm int}= {1\ov 4g^2}\sum_{m,n,p}\sum_{k,l,l'} (pk'-m'k)(nl-ml')
 \ X_{(l', n)}^a  X_{(k,p)}^a\  X_{(k',m')}^b   X_{(l,m)}^b  
}$$
+\   i{w_2\ov g} \sum_{k,m,n} m k^2  X_{2(0,m)}X^i_{(k,n)} X^i_{(-k,-m-n)} \
, 
$$
where $m'=-m-n-p ,\ \  k'=-k-l-l'$ and 
\eqn\ioi{
g^2\equiv {R_{2}^2\ov \a' }=4\pi^2R_2^3T_3\ ,\ \ \ \ \ \ 
\ww =\sqrt { w_{2}^2 k^2 + w_{1}^2 m^2 \tau_2^2 }\ ,\ \ \ \ \ \ 
\tau _2= {R_{1}\ov R_{2} }
 \ . } 
 $g$ has the interpretation of 
 type IIA string coupling. 
The operator $\a'  H_0$ is the generator of translations 
in the  world-volume time $\tau $. For the center-of-mass coordinate $X_{(0,0)}^a$ 
one obtains the simple time dependence
$X_{(0,0)}^a (\tau )= x^a +\a'  p^a \tau \ .$


The system  described by \enmo,\nome\
 can be represented as a direct product 
of the classical zero-mode  system  (which is  known 
explicitly   for all  values of $R_1,R_2$) 
 and a complicated interacting system of 
oscillation modes.
The  latter can be solved exactly 
in the special limit
 $R_{1}, R_{2}\to  \infty, T_3 \to 0$ with $\a'  $ 
 and $\tau _2$ fixed since then the interaction term  drops out.\foot{Note that if, 
  instead, we take  the limit $R_{2}\to \infty$ with $R_{1}$ fixed 
(or vice versa), then
 flat directions  remain for the constant modes in $\s $
 (or for the constant modes in $\rho )$. The system is then equivalent to the
one discussed in ref.~\russo , with an infinite number of zero modes.}
  In this limit $g^2\to \infty $, and the Hamiltonian
becomes that of an  infinite set of harmonic oscillators labelled by
$(k,m)$.
In this   
supersymmetric model, one may   argue that  since the {\it quantum}
 corrections to the masses of BPS states should not depend 
 on continuous parameters,   their values should thus 
 be the same as in this limiting case, i.e. in 
 the  case  when the interaction term 
 drops out.   This is what we shall assume below.

Let us  introduce the standard creation and annihilation operators
\eqn\ccrr{
X^i_\n ={1 \ov 2}  (\ww )^{-1/2 } \big( a_\n^i + i b_\n ^i
+ a_\n ^{i\dagger } + i b_\n ^{i\dagger } \big )\ , \ \ \ 
\ X^i_\mn =\big( X^i_\n \big)^\dagger \ , 
}
\eqn\pprr{
P^i_\n =-{i\ov 2} (\ww )^{1/2 } \big( a_\n^i + i b_\n ^i
- a_\n ^{i\dagger } - i b_\n ^{i\dagger } \big )\ , \ \ \ \ 
 P^i_\mn =\big( P^i_\n \big)^\dagger 
}
for $k=0,1,2,...$,\  $m=1,2,...$,\   and,  
\eqn\ddrr{
X^i_{(k,-m )} =\ha (\ww )^{-1/2 } \big( c_\n^i + i d_\n ^i
+ c_\n ^{i\dagger } + i d_\n ^{i\dagger } \big )\ ,
}
\eqn\drr{
P^i_ {(k,-m )}=-{i\ov 2} (\ww )^{1/2 } \big( c_\n^i + i d_\n ^i
- c_\n ^{i\dagger } - i d_\n ^{i\dagger } \big )\ ,
}
for $k=1,2,...$,\  $m=0,1,2,...$. 
Similar operators 
are introduced for $X_{2(0,m)}, \ P_{2(0,m)}$,
understanding that $a_{2\n }=b_{2\n }=0 $ if  $k\neq 0$.
Then the quadratic part of the  Hamiltonian becomes\foot{We are ignoring normal ordering 
constants since they will cancel out once fermionic 
contributions are incorporated.}
\eqn\hhgg{
\a'  H_0=\ha \a'  (p_i^2+p_{1}^2+ p_{2}^2 ) + {R_{1}^2 w_0^2\ov 2\a'  }
+{\cal H} \ ,
}
$$
{\cal H}=\sum _{m=1}^\infty \sum _{k=0}^\infty \ww \big[
a_\n ^{a\dagger }a_\n^a +b_\n ^{a\dagger }b_\n ^a\big]
+  \sum^\infty _{m=0}\sum _{k=1}^\infty \ww \big[c_\n ^{i\dagger }c_\n^i+
d_\n ^{i\dagger }d_\n ^i\big].
$$
 Using eq.~\ccnn\ one can
check the standard commutation rules of the form $[a , a^\dagger]=1$.
 The time dependence
of the mode operators is 
$a_\n^i (\tau )= e^{i\ww \tau }a_\n^i (0 ) , $  etc.  

It is convenient to define also another set  $(\a,\td \a)$ of mode operators 
\eqn\modos{
a_\n^a =i {\a _\n^a -\td \a _\n ^a \over \sqrt{2\ww } }\ ,\ \ \ 
b_\n^a = {\a _\n^a +\td \a _\n ^a \over \sqrt{2\ww } }\ ,\ \ \ 
}
\eqn\modo{
c_\n^i =i {\a _{(k,-m)}^i -\td \a _{(k,-m)} ^i \over \sqrt{2\ww } }\ ,\ \ \ 
d_\n^i = {\a _{(k,-m)}^i +\td \a _{(k,-m)} ^i \over \sqrt{2\ww } }\ , 
}
satisfying  
\def\www {\omega _\n }
\eqn\crul{
[ \a _\n^a , \a^b_{(k',m')}]= \www \delta _{k+k'}\delta _{m+m'}\delta^{ab}
\ ,\ \ \  \ \ \www =  \epsilon (k ) \ \ww  \ , 
}
where $\epsilon (k )$ is the sign function,  and similar relations 
 for the $\tilde \a _\n ^a$.
 Then the oscillator
part of the Hamiltonian becomes 
\eqn\inte {
{\cal H}=\ha \sum _{m,k}^\infty \big[ \a _\mn^a\a _\n^a +
\td\a _\mn^a\td\a _\n^a\big]\ .
}
Explicitly, the time dependence of $X^a$ is as follows
$$
X^a = x^a +\a'  p^a \tau +
i \sqrt{\textstyle {\a' \ov 2}} \sum_{\n \neq (0,0)}
 \www\inv  \big[ \a _\n^a e^{ik\s +im\rho }  
+ \td \a _\n ^a e^{-ik\s -im\rho }
\big] \ e^{i\www \tau } .
$$


The requirement that $X^-$ is single-valued  leads to the following
  global constraints for the physical states (see e.g. 
  \refs {\dewitt ,\dufi, \russo })
\eqn\vvs{
{\bf P^{(\s )} }={1\ov 2\pi \a' } 
\int _0^{2\pi } d\s \ \del_\s X^a \dot X^a \equiv 0\ , \ \ 
\ {\bf P^{(\rho )} }= {1\ov 2\pi \a' }
\int _0^{2\pi } d\rho \ \del_\rho X^a \dot X^a \equiv 0\ .
}
The operators $ {\bf P^{(\s )} },\ {\bf P^{(\rho )} } $ generate
translations in $\s $ and $\rho $.
These conditions were discussed in ref.~\russo\ for the space
\rs . It is easy to generalize the expressions 
for the present case. Let the momenta in the directions $X^{1}$ and $X^{2}$
be given by
$$
p_{1}={l_1\ov R_{1} }\ ,\ \ \ \ p_{2}={l_2\ov R_{2} }\ ,\ \ \ \ 
\ l_1, l_2 \in {\bf Z} \ . 
$$
Inserting the expansions for $X^a$ in terms of the mode operators 
we find\foot{There are changes in the  notation relative to ref.~\russo.
Now $N_\s^+, N_\s^-, N_\rho^+, N_\rho^-$ stand for
${\bf N}, {\bf \td N}, {\bf N^+},{\bf N^-}$, and $\td \a _{(k,-m)}\to
\td \a _{(k,m)}$. Also, in \russo , $w_{1}=0$, $w_{2}=1$, $Q=l_2$.}
\eqn\lvmv{
N^+_\s -N^-_\s =w_{1} l_1\ , \ \ \ \ \ \ \ \ N^+_\rho -N^-_\rho =w_{2} l_2\ ,
}
where
\eqn\zzxx{
N^+_\s = \sum _{m=-\infty }^\infty \sum _{k=1}^\infty {k\ov \ww }
\a^i_\mn \a^i_\n\ , \ \ \ \  N^-_\s = \sum _{m=-\infty }^\infty \sum _{k=1}^\infty {k\ov \ww }
\td \a^i_\mn \td \a^i_\n\ ,
}
\eqn\ttxxz{
N^+_\rho=\sum _{m=1}^\infty \sum _{k=0}^\infty {m\ov \ww }
\big[ \a^a_\mn \a^a_\n + \td \a^a_{(-k,m)} \td \a^a_{(k,-m)} \big]\ ,
}
\eqn\ttxz{
N^-_\rho=\sum _{m=1}^\infty \sum _{k=0}^\infty {m\ov \ww }
\big[ \a^a_ {(-k,m)}\a^a_{(k,-m)} + \td \a^a_{(-k,-m)} 
\td \a^a_{(k,m)} \big]\ .
}
%
%
%
As usual, the    Fock vacuum $|0\rangle $ is defined 
as the state annihilated by the $a^a_\n ,$ $  b^a_\n ,$ $  c^i_\n ,$ $ 
 d^i_\n $,
and $p_a|0\rangle =0$, and  the Fock space is generated by the states 
constructed by successive applications of the creation operators
to  the vacuum.
The  physical Hilbert  space  thus  consists  of all states in the 
Fock space obeying the conditions \lvmv.

\subsec{\bf Matching the membrane and   type IIB string  BPS spectra}

The nine-dimensional  membrane mass operator is given by
\eqn\madd{
M^2=2p^+ p^- -  p_i^2 = 2H_0  - p_i^2
= {l_1^2\ov R_{1}^2} + {l_2^2\ov R_{2}^2}+ {w_0^2 R_{1}^2\ov {\a' } ^2}
+ {2\ov \a'  } {\cal H}\ ,
}
where ${\cal H}$ is given  by \inte .
Our aim is  now  to compare  the  BPS part of this 
membrane   spectrum    with the 
type IIB  $(q_1,q_2)$ string BPS spectrum 
given in  \john, where  the correspondence
 between the  spectra was established
at the zero-mode level. We shall show that it extends  to the oscillator
level as well.

{}From the ten-dimensional string-theory point of view, the  Kaluza-Klein momentum  
 $p_{2}=l_2/R_{2}$   corresponds to 
the  Ramond-Ramond charge. Let us  first consider 
 the perturbative  or $(1,0)$ string states with    $l_2=0$, the   
 winding number $l_1$, the Kaluza-Klein momentum $w_0/R'_{1}$
 and the mass 
(note  that under $T$-duality relating IIA and IIB spectra 
$R_{1} \to R_1'=\a'  /R_{1}$) 
 \eqn\dosb{
 M_{\rm IIB}^2= {l_1^2 {R_{1}'}^2\ov {\a' }^2 } + {w_0^2\ov {R_{1}'}^2}
 +{2\ov \a'  } (N_R +N_L) \ , \ \ \ \ \ N_R-N_L=l_1 w_0 
\ ,
} 
and compare  general BPS  states  (having $N_L=0$ or $N_R=0$)
with the  corresponding ($l_2=0$) 
oscillating states  in the membrane spectrum \madd.
For a BPS saturated state, the mass  should take  the minimal  possible
value compatible with the charges. 
For the membrane, these are the states with
$N^-_\s =N^-_\rho =0$. The constraint equations \lvmv\ 
become
\eqn\otra{
N^+_\s = l_1 w_{1} \ ,\ \ \ \ \ \  \ \ N^+_\rho =0 \ .
}
The condition $N^+_\rho =0$ implies that such states are  constructed  by
applications of $\a^i_{(-k,0)} $ 
to  the vacuum. For these states, $\omega _{k0}=w_2 k$, and
(cf. \inte, \zzxx ) $ \ 
{\cal H}= w_{2} N_\s ^+ =w_1 w_2 l_1, $ 
so that
\eqn\mmmb{
M^2 = {l_1^2\ov R_{1}^2} + {w_0^2 R_{1}^2\ov {\a' } ^2}
+{2\ov \a' } l_1 w_{1} w_{2}= \bigg({l_1\ov R_{1}} +{w_0 R_{1}\ov {\a' } }\bigg)^2
\ .
}
This coincides with  the masses of perturbative  string BPS states 
 obtained upon setting $N_L=0$ in \dosb.


 The mass formula for the  free string  states of 
 the  non-perturbative type IIB   string  with charges $(q_1, q_2)$
   given in \john\ is 
$$
M_{\rm IIB}^2= \big(2\pi R_1' n T_{(q_1,q_2)}\big)^2 +
{w_0^2 \ov {R'_{1}}^2 }+ 4\pi T_{(q_1,q_2)} (N_L + N_R)
$$
\eqn\mmbb{
= {l_1^2 {R'_{1}}^2 \ov {\a' }^2} +
 {l_2^2\ov R_{2}^2 }+{w_0^2 \ov {R'_{1}}^2 } 
+ 4\pi T_{(q_1,q_2)} (N_L + N_R)\ , 
}
\eqn\llvv{  \ N_R-N_L=n w_0\ , }
where $l_1=n q_1 ,\  l_2=n q_2 $, with $q_1, q_2 $  being co-prime, and 
\eqn\tensi{
T_{(q_1,q_2)}= {T_2\ov n} \sqrt {l_1^2 + \tau _2 ^2 l_2^2 } \ ,\ \ \ \ 
T_2=(2\pi \a' )^{-1}\ ,\ \ \ \tau_2  = 
{R_{1}\ov R_{2} } = e^{-\phi_0} 
\ .
}
Eq. \mmbb\  should  be exact for BPS states \john. 
The case of NS-NS string \mmbb\  corresponds to 
 $l_2=0,  q_1=1,q_2=0$. In the case of the
   R-R string  $l_1=0$, $q_1=0,q_2=1$ 
  one finds that for BPS states with $N_L=0$  
\eqn\zzzs{
M_{\rm IIB}^2={l_2^2\ov R_{2}^2 }+ {w_0^2 R_{1}^2\ov {\a' } ^2}
+ 4\pi T_{(0,1)} \big( N_L + N_R \big) = 
\bigg( {l_2\ov R_{2} } +
{w_0 R_{1}\ov {\a' } }\bigg)^2\  .
}
In the supermembrane spectrum \madd\  the minimal  mass for given charges
is obtained for states
with $N_\rho ^-=N_\s ^-=0$.
For the counterpart of the   R-R states with $l_1=0$,
 the constraints \lvmv \   are similar to \otra\ 
$
N_\s ^+= 0 ,\  N^+_\rho  = l_2 w_{2} .\
$
$N_\s ^+$ is a sum of positive definite terms, so 
it is equal to zero only if each term in it vanishes.
The most general physical state satisfying $N_\s ^+= N_\s ^- =0$ 
is thus obtained by acting  by 
 $\a _{(0, -m )}^i $  on  the vacuum.
On these states, $N_\rho ^+$  and  ${\cal H}$ take 
 the form (see \inte)
\eqn\nnpp{
N^+_\rho = {1\ov w_{1} \tau_2}
\sum _{m=1}^\infty  
\a _{(0,-m)}^a \a _{(0,m)}^a \ ,  \ \ \ \ 
{\cal H}= w_{1} \tau _2 N_\rho ^+ =l_2 w_0  \tau _2\ , 
}
so that 
\eqn\abcd{
M^2= {l_2^2\ov R_{2}^2 }+ {w_0^2 R_{1}^2\ov {\a' }^2 }
+ {2\ov \a' } \tau _2  l_2 w_0=\bigg( {l_2\ov R_{2} } +
{w_0 R_{1}\ov \a'  } \bigg)^2 \ ,
}
 in  agreement with  the type IIB string expression \zzzs . 

The previous two examples of comparison 
with the 
$(1,0)$ and $(0,1)$ string spectra 
illuminate  the fact that the  BPS states 
in the
supermembrane spectrum  have 
excitations  only  in one direction -- along the momentum vector.
This  is   also  what is suggested   by   correspondence between 
the $(q_1,q_2)$  string  and wrapped membrane   with momentum along
$(q_1,q_2)$ cycle
at the level 
of the classical 
solutions  discussed in  Section 2. 
Let us now identify the  relevant  oscillation  states in the 
supermembrane spectrum for generic  $(q_1,q_2)$.
Performing  the  rotation as in \rott,\sini,   
$
y_1= \cos\theta \ X_{1} + \sin \theta\  X_{2} ,\ \ 
y_2= -\sin\theta \ X_{1} + \cos \theta \ X_{2},  
$
  where, for generic  $R_1,R_2$, 
   $q_2 \to {R_1\ov R_2} q_2 $, i.e.   
$$
\tan \theta = { q_2R_1\ov q_1R_2}\ ,
$$
we may align the momentum  with  the direction $y_1$.
The map between the target-space  torus and the toroidal 
membrane surface  is given by the zero-mode part
\eqn\zeee{
y_1^0= w_{1} R_{1} \cos \theta \ \s  + w_{2} R_{2} \sin \theta \ 
\rho\  \ , 
\ \ \ \ \ 
y_2^0= -w_{1} R_{1} \sin \theta \ \s  + w_{2} R_{2} \cos \theta
 \ \rho  \ . 
}
For an  oscillation mode $\a^i_{(k,m)}e^{i (k\s +m \rho )}$
the  factor in the exponent can be written as  
$$k\s + m\rho = \bigg( {k\ov w_{1} R_{1}} \cos\theta + {m\ov w_{2} R_{2}}
\sin\theta \bigg) y^0_1 + \bigg( -{k\ov w_{1} R_{1}}
 \sin\theta + {m\ov w_{2} R_{2}}
\cos\theta \bigg) y^0_2 , $$
so that   the condition that there are no oscillations along $y_2$ 
becomes 
$
\a ^i_{(k,m)}=0$ if 
$w_{2}R_{2} k \sin \theta \neq w_{1} R_{1} m \cos \theta $, 
i.e.  if $w_2 k q_2 \not= w_1 m q_1$. 
 Thus the relevant states are
constructed    using  $\a _{(-k, -m_0)}^i $  with 
\eqn\ccff{   
m_0=   {w_{2}  q_2 \ov  w_{1} q_1   } k \ .
}
For  such  states,
$${\cal H}=\sum _{k=1}^\infty \a^i_{(-k,-m_0)}\a^i_{(k,m_0)}\ , \ \ \ \ \
\ \ 
 \omega_{km_0}  = {kw_{2}\ov q_1}\sqrt{  q_1^2 +  q_2^2 \tau _2^2 } \ ,
$$
and the constraints become 
(understanding that $\a^i_{(k,m_0)}=0 $ if $m_0\not \in {\bf Z}$)
\eqn\conrr{
N_\s^+={q_1\ov  w_{2}\sqrt{ q_1^2+ 
 q_2^2 \tau _2^2 } } \sum _{k=1}^\infty  \a^i_{(-k,-m_0)}\a^i_{(k,m_0)}
=l_1 w_{1} \ ,
}
\eqn\crrr{
N_\rho^+={q_2\ov   w_{1} \sqrt{ q_1^2+ q_2^2 \tau _2^2 } } 
\sum _{k=1}^\infty 
 \a^i_{(-k,-m_0)}\a^i_{(k,m_0)}
=l_2  w_{2} \ . 
}
The  membrane  BPS mass formula is then 
\eqn\sssq{
M^2 = {l_1^2\ov R_{1}^2 }+ {l_2^2\ov R_{2}^2 }+
{w_0^2 R_{1}^2\ov {\a' }^2 } + {2{\cal H}\ov \a' }\ ,
}
with
\eqn\hhc{
{\cal H}= {w_{2}\ov q_1}  \sqrt{  q_1^2 +  q_2^2 \tau _2^2 }\ 
N_\s ^+ \ = \ 
w_0\sqrt{ l_1^2 +l_2^2 \tau _2^2} \ .
}
Using \conrr , \sssq\ and \hhc , we thus  obtain
\eqn\rmk{
M^2=\bigg( \sqrt { {l_1^2\ov R_{1}^2} +{l_2^2\ov R_{2}^2 } }
+ {w_0 R_{1}\ov \a'  } \bigg)^2 \ .
}
Remarkably, this agrees  with the Schwarz string 
 mass formula \mmbb , \llvv\  for BPS states 
with $N_L=0$.

\newsec{Acknowledgements}
A.A.T.  
would like to thank  G. Papadopoulos  and P. Townsend 
for useful  correspondence and also 
 acknowledges
 the support of PPARC and 
 the European
Commission TMR programme ERBFMRX-CT96-0045.
\vfill\eject
\listrefs
\end